\documentclass[twoside]{article}

\usepackage[sc]{mathpazo} 
\usepackage[T1]{fontenc} 
\usepackage{microtype} 

\usepackage[hmarginratio=1:1,top=32mm]{geometry} 

\usepackage[french,english]{babel}

\usepackage{amsmath,amsfonts,amssymb}
\usepackage{graphicx}

\usepackage{fancyhdr} 
\usepackage[latin1]{inputenc}

\def\arcsec{{\tt "}}

\begin{document}
 
\title{\vspace{-15mm}\fontsize{20pt}{10pt}\selectfont\textbf{
PISCO2: the new speckle camera of the Nice 76-cm refractor}
}

\author{ R. Gili$^1$, 
J.-L. Prieur$^{2,3}$\thanks{Corresponding author:
jean-louis.prieur@irap.omp.eu},
J.-P. Rivet$^4$, 
F. Vakili$^{1}$, 
L. Koechlin$^{2,3}$ \\
D. Bonneau$^4$ 
\\
\\
$^1$ 
UMS 2202, Universit\'e de Nice Sophia Antipolis -- CNRS\\
Observatoire de la C\^ ote d'Azur,
CS 34229, 06304 Nice Cedex 4, France \\
$^2$ CNRS -- IRAP, 14 avenue E. Belin, 31400 Toulouse, France \\ 
$^3$ Universit\'e de Toulouse -- UPS-OMP -- IRAP, Toulouse, France \\ 
$^4$ 
UMR 7293, Universit\'e de Nice Sophia Antipolis -- CNRS\\
Observatoire de la C\^ ote d'Azur,
CS 34229, 06304 Nice Cedex 4, France \\
}

\tolerance=4000

\thispagestyle{fancy} 


\date{Received \today; accepted }

\maketitle 

\begin{abstract}
We present the new speckle camera PISCO2 
made in 2010--2012, for the 76-cm refractor of C\^ote d'Azur Observatory. 
It is a focal instrument dedicated to the observation of 
visual binary stars using high angular resolution 
speckle interferometry techniques to partly overcome the degradation
caused by the atmospheric turbulence. 
Fitted with an EMCCD detector, PISCO2 allows the
acquisition of short exposure images that are processed in real
time by our specially designed software. Two Risley prisms
are used for correcting the atmospheric dispersion. All optical settings
are remotely controlled. 
We have already been able to observe
faint close binary stars
with angular separations as small as 0\arcsec.16, and visual magnitudes
of about~16. We also have measured some particularly difficult
systems with a magnitude difference between the two components
of about~4~mag.
Those performances are very promising for the detection and study
of large sets of yet unknown (or partly measured) binaries
with close separation and/or large magnitude difference.
\end{abstract} 
 
\tolerance=4000

\section{Introduction}
\label{sec:intro}

This paper presents the new speckle camera PISCO2 
(Pupil Interferometry Speckle camera and COronagraph, 2nd version) 
made in 2010--12 for the 76-cm refractor
telescope (``Grand Equatorial de l'Observatoire de la C\^ote d'Azur'',
hereafter L76).
PISCO2 is a focal instrument the purpose of which is to provide 
high angular resolution images using speckle interferometry techniques.
Those techniques allow to partly overcome the degradation
caused by atmospheric turbulence (Labeyrie, 1970). 

PISCO2 is a simplified version of PISCO 
that was developed
in 1993 for the 2-meter Bernard Lyot telescope (Pic du Midi, France). 
PISCO is a multi-purpose focal instrument 
with many observing modes: pupil interferometry,
pupil-mask aperture synthesis, SCIDAR atmospheric turbulence measurements,
grism spectroscopy, coronagraphy, and Shack-Hartmann wavefront sensing.
A detailed presentation can be found in Prieur et al.~(1998).
Since 2004 PISCO has been operated 
on a dedicated 1-meter telescope in Merate (Brera Observatory, Italy) 
(see e.g., Scardia et al., 2007, 2013).

The optical design of PISCO2 is similar to that of PISCO but the 
observing modes are reduced, and are mainly limited
to speckle observations.
Indeed, PISCO2 was specially designed for the observation of 
visual binary stars. Associated with an electronic detector, it allows the
acquisition of enlarged short exposure images exhibiting
``speckles''. Those images can then
be processed as first suggested by Labeyrie~(1970), to provide
high-angular resolution information. 

In this paper, we first present the L76 refractor (Sect.~\ref{sec:L76})
and the modifications we have done to make it fully operational.
The instrumental setup of PISCO2 
and its main technical capabilities 
are described in Sect.~\ref{sec:pisco2-presentation}. 
The atmospheric-dispersion corrector
which is using Risley prisms is presented in Sect.~\ref{sec:risley}. 
The absolute scale calibration of PISCO2 
can be done with a grating mask placed 
on the entrance pupil of the telescope. The whole calibration procedure is 
explained in Sect.~\ref{sec:calibration}.

\begin{figure}
\centerline{
\includegraphics*[height=7.5cm]{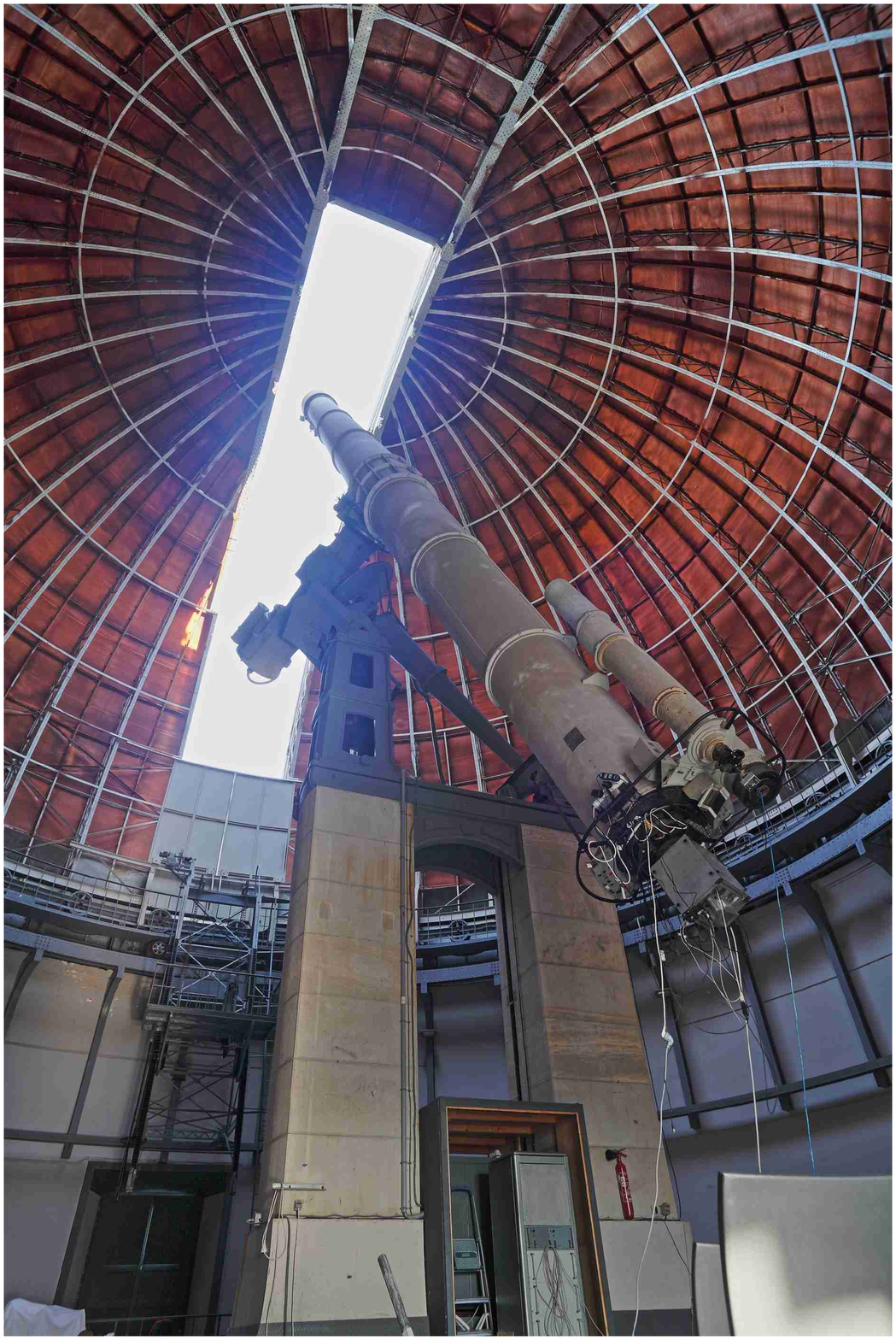} 
\includegraphics*[height=7.5cm]{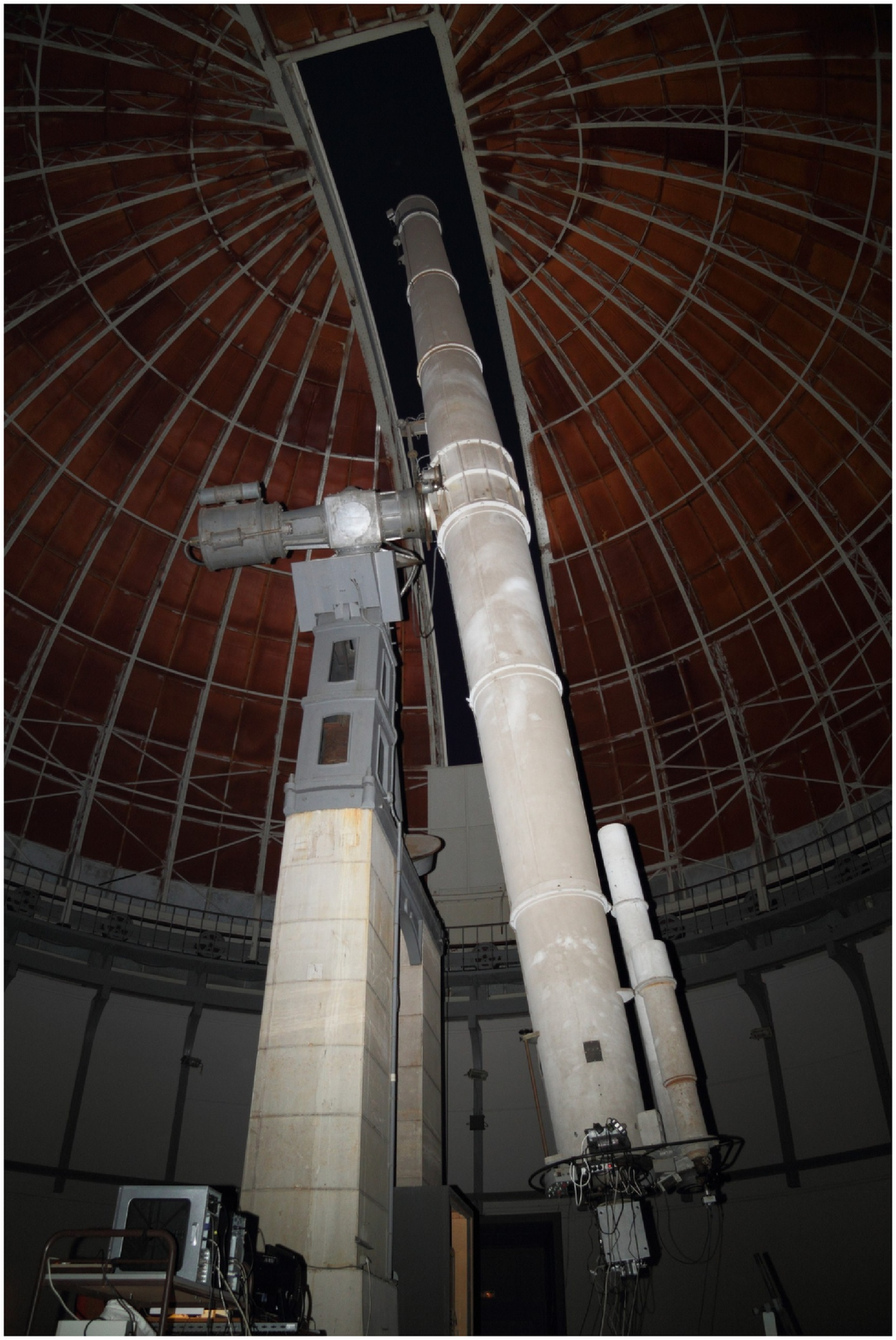} 
\includegraphics*[height=7.5cm]{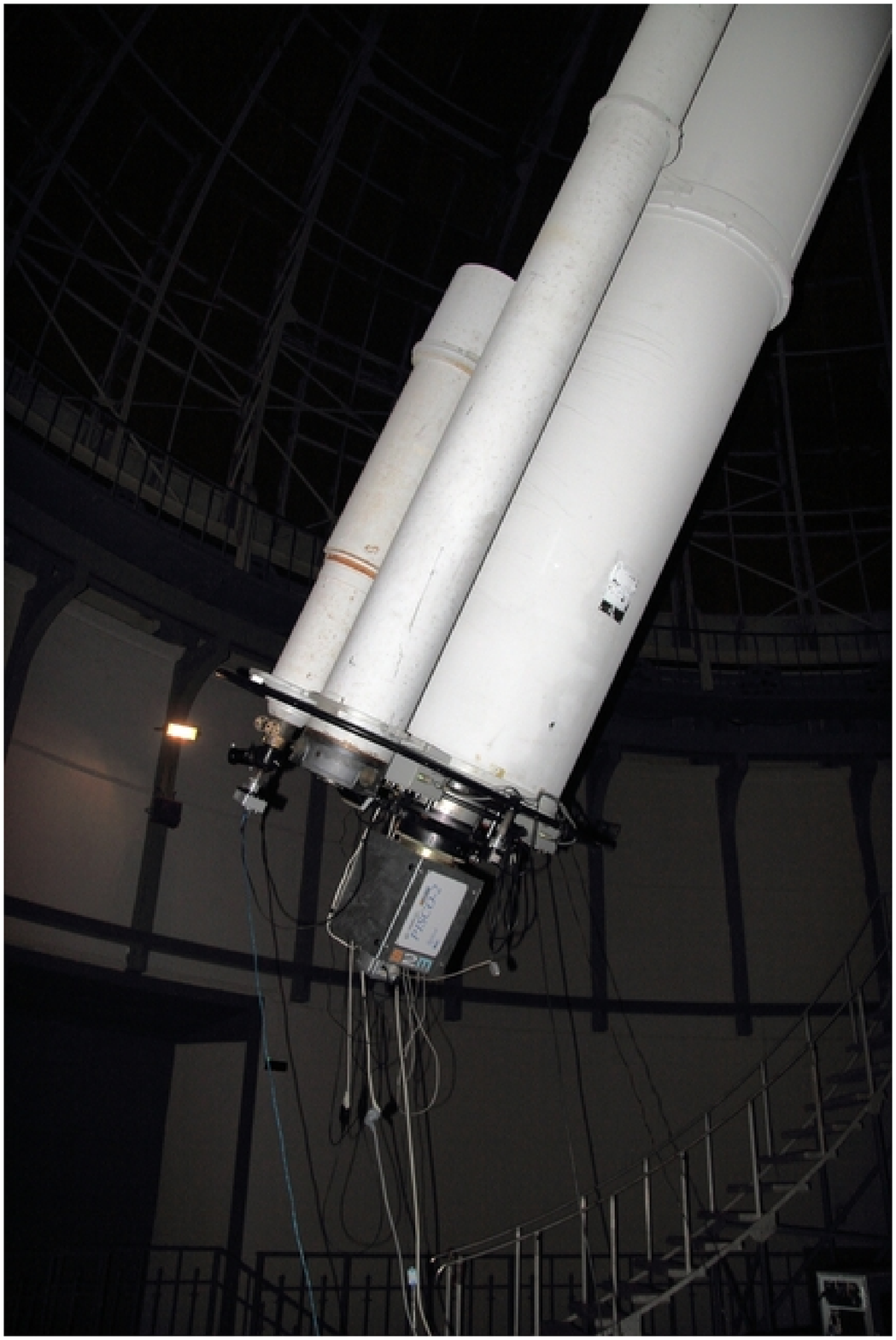} 
}
\caption{PISCO2 on the Nice 76-cm refractor (L76). The 
largest finder which is mounted on ``piggy-back'' is
a Zeiss 25-cm refractor. 
}
\label{fig:L76}
\end{figure}

\begin{figure*}
\centerline{
\includegraphics*[height=6cm]{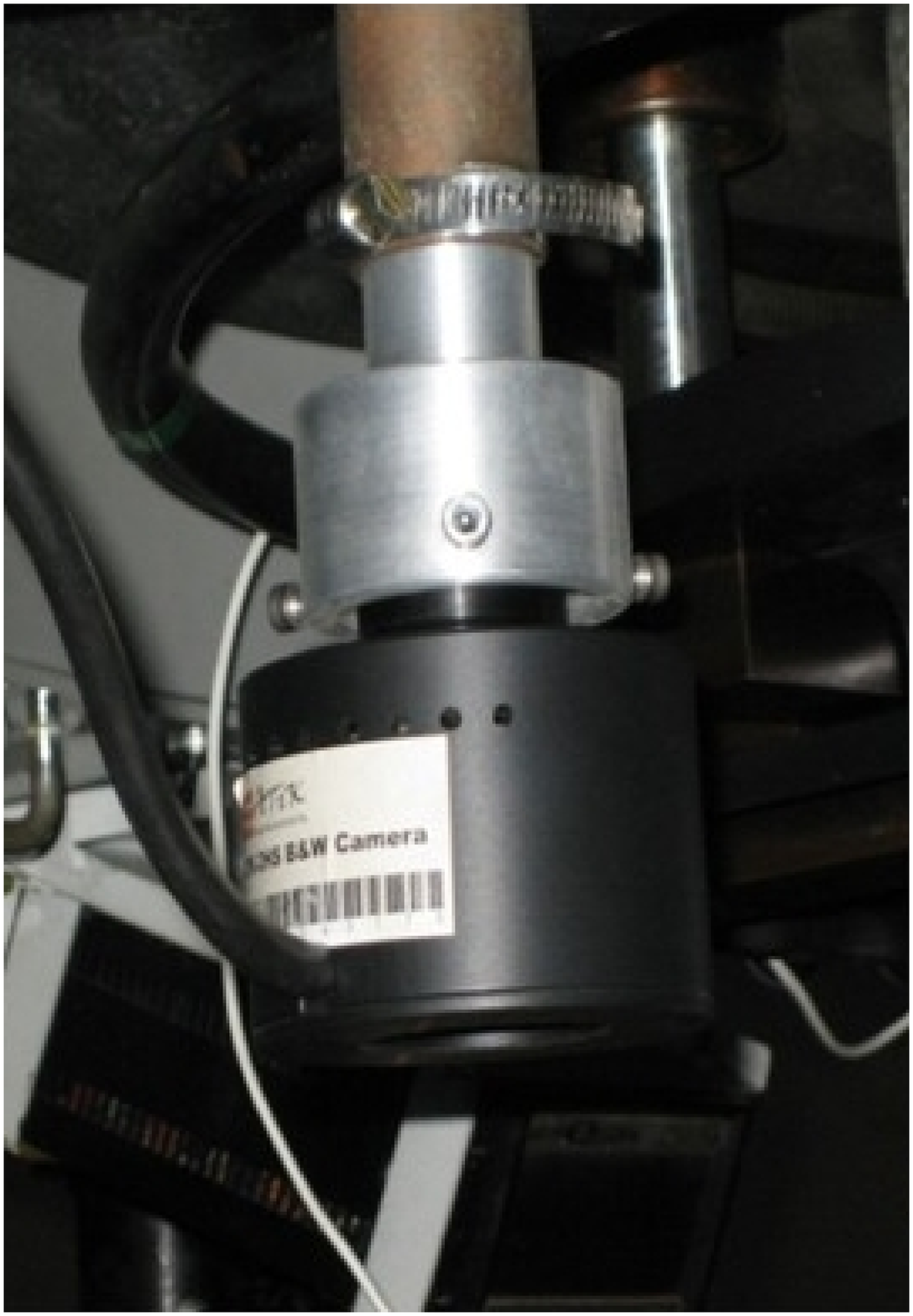}
\includegraphics*[height=6cm]{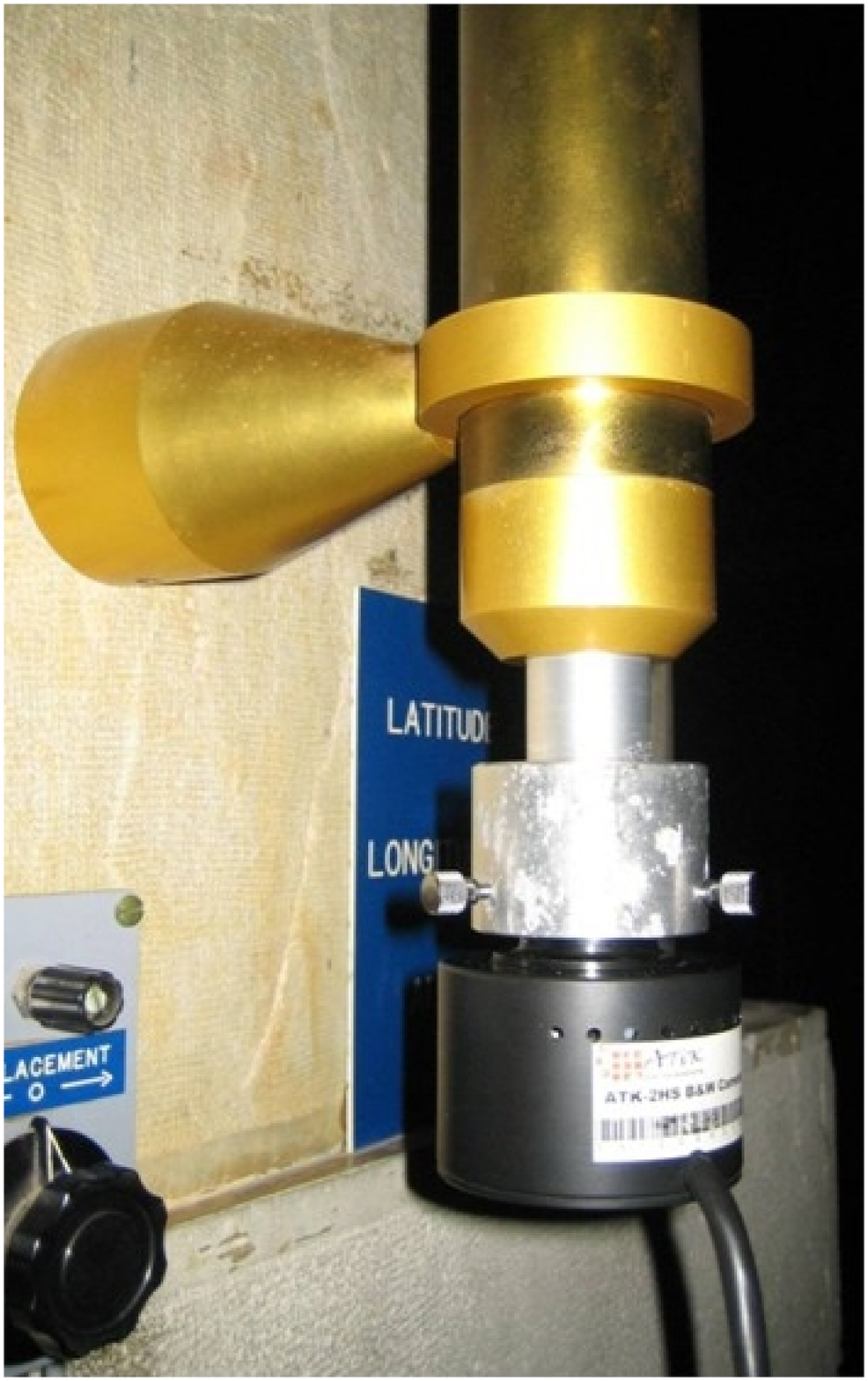}
\hskip-3.5mm
\vbox to 6cm{
\vglue 1mm
\hbox{
\begin{tabular}{c}
\includegraphics*[height=4.6cm]{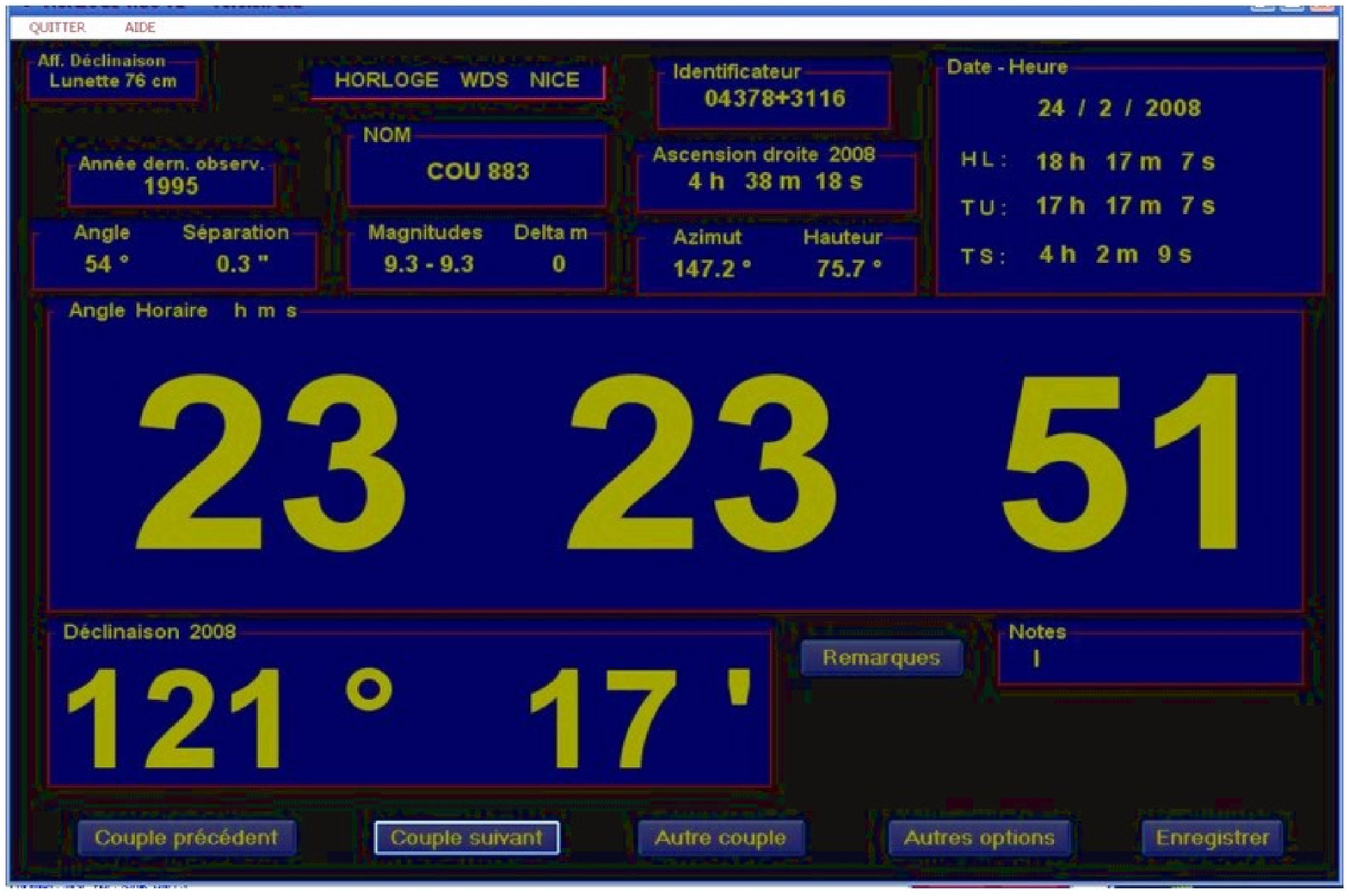} \\
\hbox{
\includegraphics*[height=1.2cm]{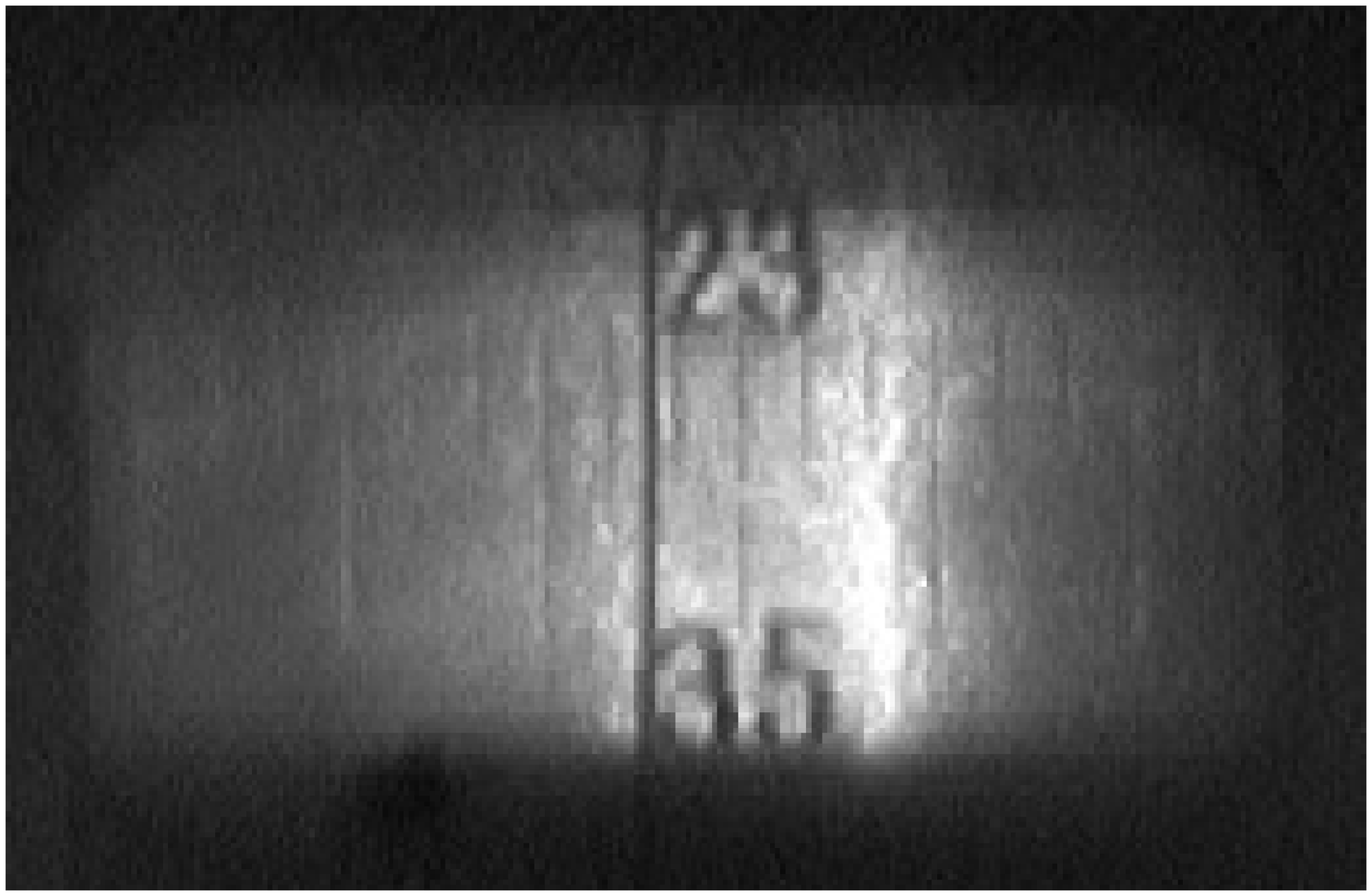} 
\includegraphics*[height=1.2cm]{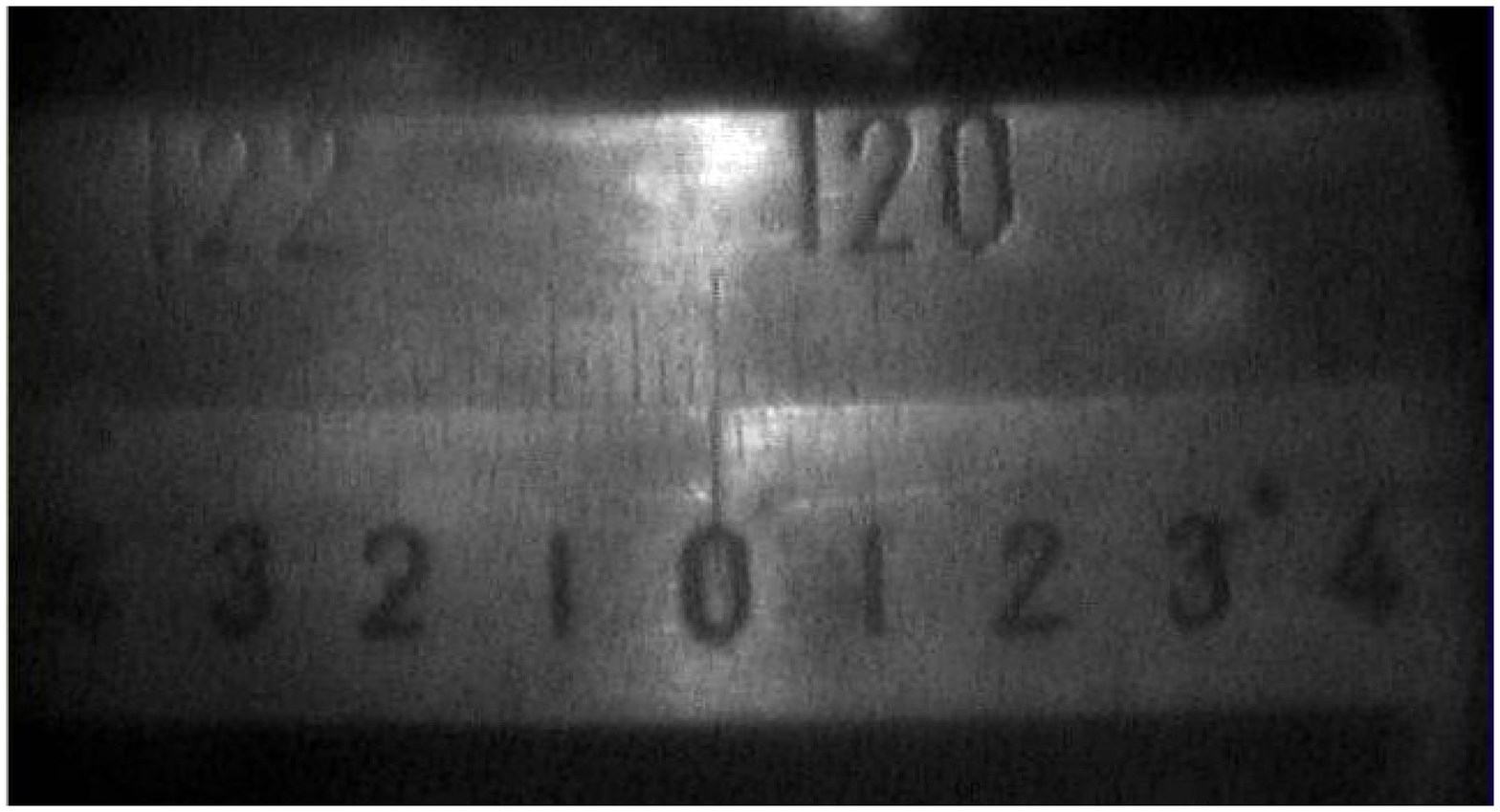}
}
\end{tabular}
}
\vfil
}
}
\caption{Webcams used for reading the hour angle (H.A.) and declination
(Dec.) circles and thus allowing remote control of the L76 refractor.
Left panel: H.A. webcam, central panel: Dec. webcam, 
upper right panel: \texttt{Horloge} program
(from G. Morlet), lower right panel: 
webcam images of the H.A. and Dec. circles.}
\label{fig:webcams}
\end{figure*}

\begin{figure*}
\centerline{
\includegraphics*[height=6cm]{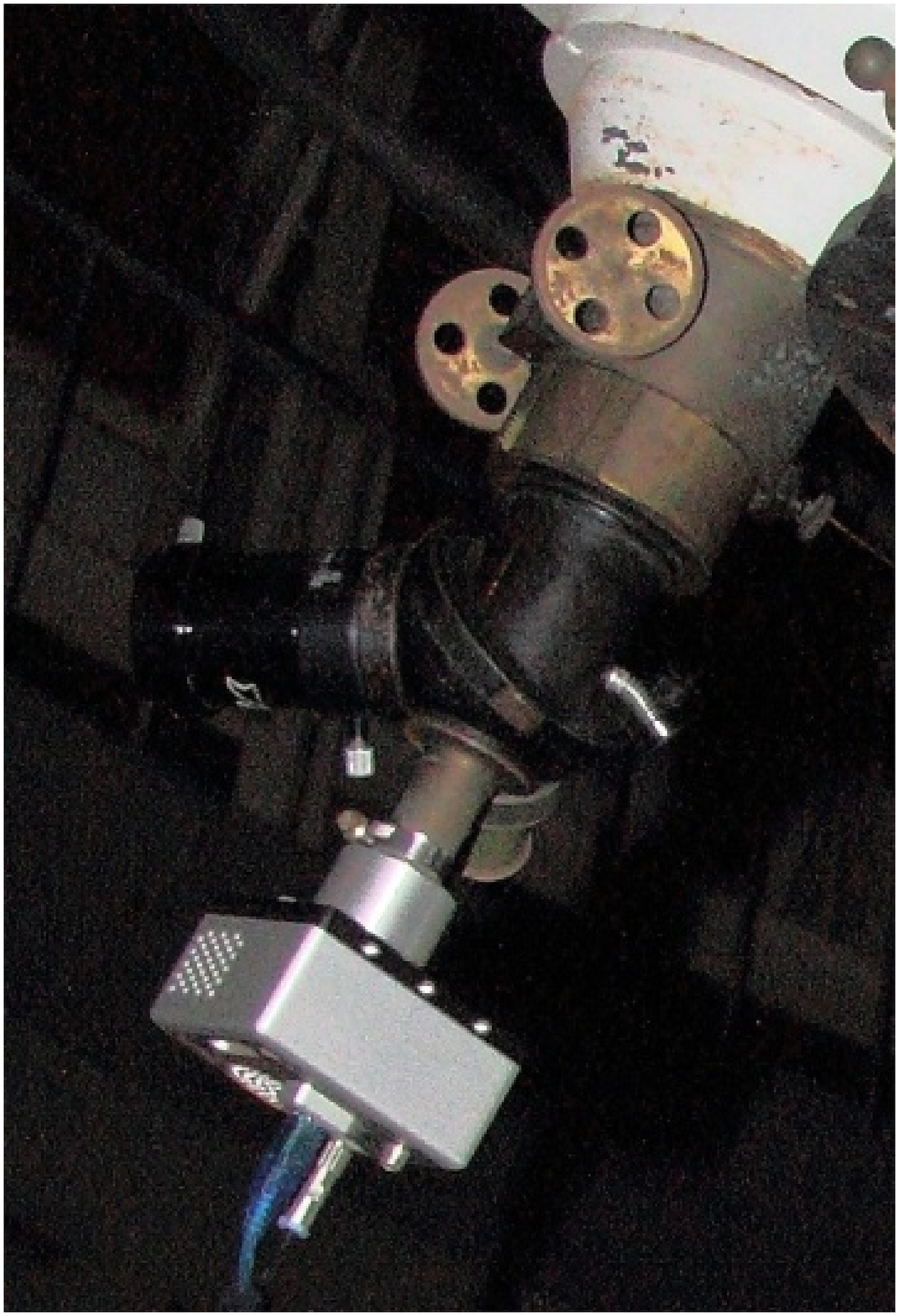}
\includegraphics*[height=6cm]{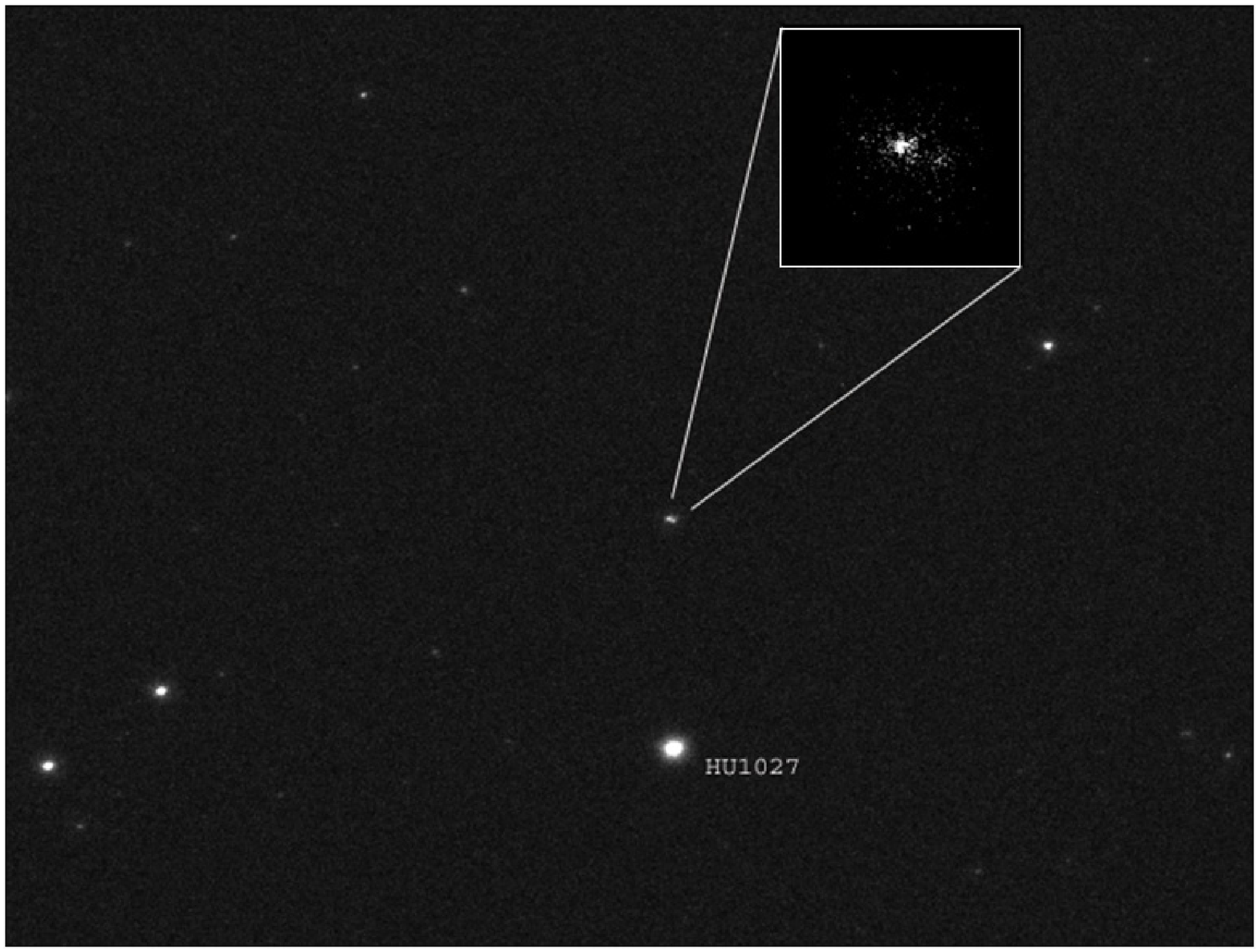}
}
\caption{ANDOR LUCAs camera at the focus of the Zeiss refractor (left)
which is used as a wide-field finder for speckle observations
with the L76 refractor. Right: comparison of the corresponding
field of view (9'$\times$12')
with that provided by the ANDOR DV885 camera on the L76 (16\arcsec$\times$16\arcsec).}
\label{fig:luca-zeiss}
\end{figure*}

\section{The Nice 76-cm refractor} 
\label{sec:L76}

PISCO2 was specially designed for the
Nice 76-cm refractor 
belonging to Observatoire de la C\^ote d'Azur~(OCA).
Built in 1887, when it was the largest refractor in the world,
this famous telescope has a free aperture diameter 
of $D=74$~cm and a focal length of 17.89~m
(see Fig.~\ref{fig:L76}).
It is one of the largest refractors still in operation in the world. 
It has mainly been devoted to binary observation
since its construction.
The good quality of its optics makes Airy rings clearly visible when the seeing
is good. And this happens very frequently, despite the vicinity 
to Nice city. 

Paul Couteau is probably the most famous astronomer who has 
used it. He was the architect of the renovation of this refractor 
from 1965 to 1969 (Couteau, 1970) and many 
binaries of his impressive Couteau's catalogue 
(Couteau, 1993), were measured with this telescope 
(e.g., Couteau \& Gili, 1994, Gili \& Couteau, 1997). 
When he retired in 2000, one
of his closest collaborators, R. Gili (hereafter RG),
decided to continue this long story of binary observations
with the L76. Although he was an experimented binary star observer
with a filar micrometer, RG wanted to use more modern techniques,
like precise binary measurements on short-exposured CCD 
(Charge-Coupled Device) images that he had already obtained in collaboration
with other colleagues (Salaman et al., 1999, Morlet et al., 1999, 2002,
Gili \& Bonneau, 2001).
For this new operating mode, it soon appeared 
that the L76 needed a serious revision
and that a specially designed
speckle camera was needed for improving the efficiency and
quality of those observations.
The 76-cm lens was then dismounted, thoroughly cleaned, and
the two (crown and flint) components were fitted into a new 
improved mechanical mount. The dome motorisation was thoroughly 
revised, both for the electronics and the mechanics.

In 2008, two webcams were fitted to the existing magnifying
optics of the
hour angle and declination circles to allow 
their reading from a dedicated control computer (Fig.~\ref{fig:webcams}). 
This allowed the observer to safely control the telescope
while remaining sitted in front of a control computer.
Before this modification, for
pointing any object with the telescope,
the observer had to move to each of the two coordinates circles 
and then climb on top of a ladder to watch the star 
in the finder. Obviously, this could be dangerous in the darkness,
especially when the observer started to be tired and feeling sleepy (!).
Due to the large size of the telescope (about 18~m),
this was very particularly dangerous when the telescope was 
pointing to low elevation targets. 
A few months later, 
a new ANDOR DV885 camera was acquired for speckle
observations, and replaced the ANDOR LUCAs that was formerly used for this task.
This ANDOR LUCAs camera was installed on the Zeiss 25~cm 
finder, and was thus able to transmit field images to the control
computer (see Fig:~\ref{fig:luca-zeiss}). 
This happened to be a decisive improvement to the pointing procedure,
since from then both the L76 telescope
and its focal instrumentation could be operated from this computer, by a single
observer.

Focusing the telescope was another hazardous operation, especially
for observations done by a single observer (by far the most frequent case).
For CCD or EMCCD observations, the observer had to 
climb on the top of a large step-ladder to actuate the big focusing hand-wheel,
then go down for checking the quality of the CCD image 
on the computer screen. This had to be done a few times until
satisfactory focusing was achieved.
Furthermore, this operation had to be done in full darkness, which was 
needed to avoid saturating the camera. 

The few thousands of measurements published in Gili \& Agati (2009)
and Gili \& Prieur (2012) were obtained in those difficult conditions. 
Indeed, the original focusing system was part of the L76 mechanics
and could not be motorized easily, without a thorough modification 
of this historical instrument.
Consequently, an important specification for the design of PISCO2 
was to include a remote control focusing capability.

\section{Presentation of PISCO2} 
\label{sec:pisco2-presentation}

The decision to build a new speckle camera for the L76
was taken in 2009. The main idea was to ease and speed up the observations 
while extending the observable domain to lower elevation targets.
The basic constraints were: simplicity, lightness, small overall dimensions,
and low cost. The last constraint was rather severe since most of the
expenditure had to be covered by personal funds.
For this design, we took profit of our experience in binary star observations
with the L76 and with PISCO on other telescopes. 

\begin{figure*}
\centerline{
\includegraphics*[height=6cm]{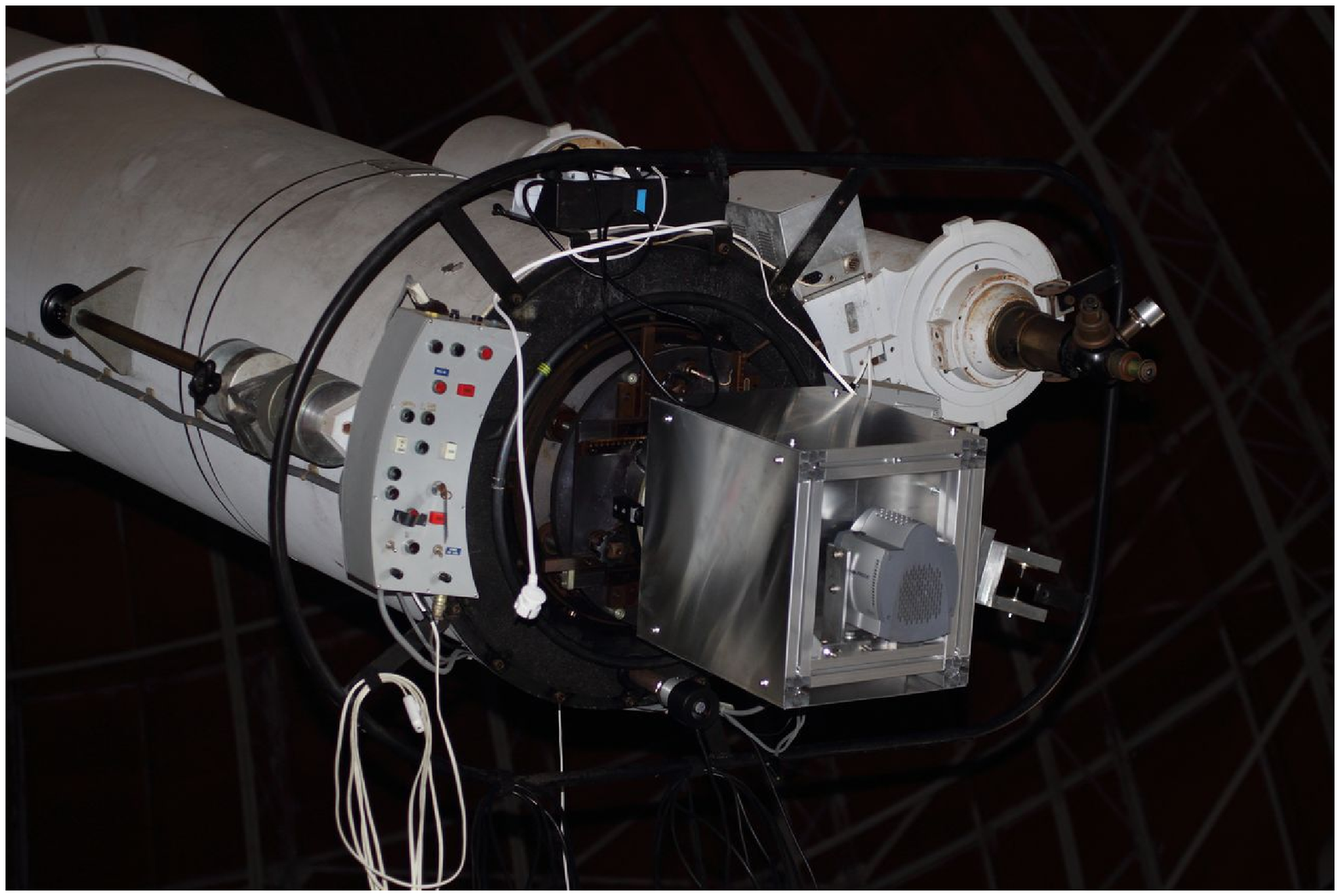} 
\includegraphics*[height=6cm]{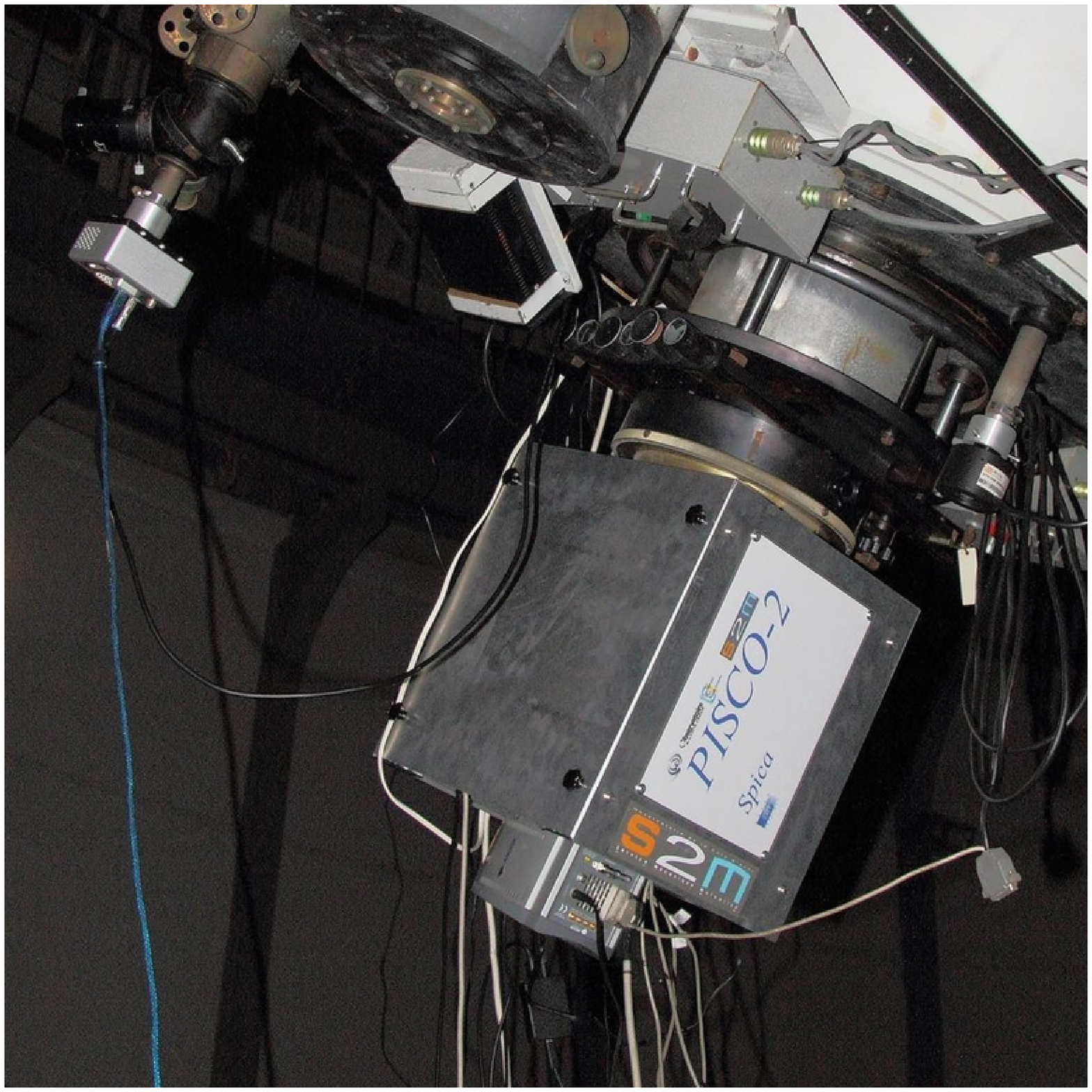} 
}
\caption{PISCO2 with the ANDOR DV897 EMCCD camera
at the focus of the 76-cm refractor.
}
\label{fig:pisco2-L76}
\end{figure*}

\begin{figure*}
\centerline{
\includegraphics*[height=6cm]{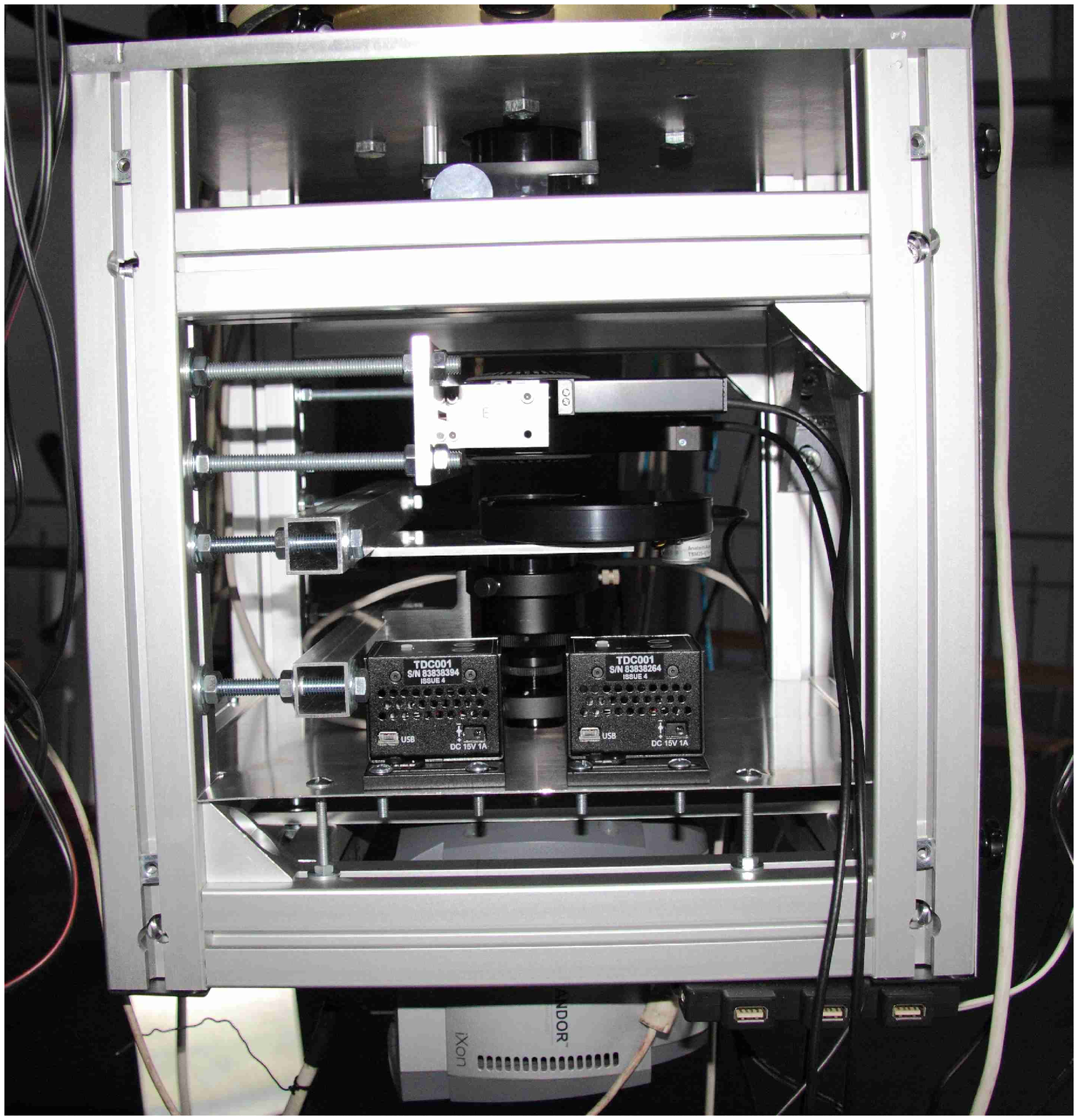} 
\includegraphics*[height=6cm]{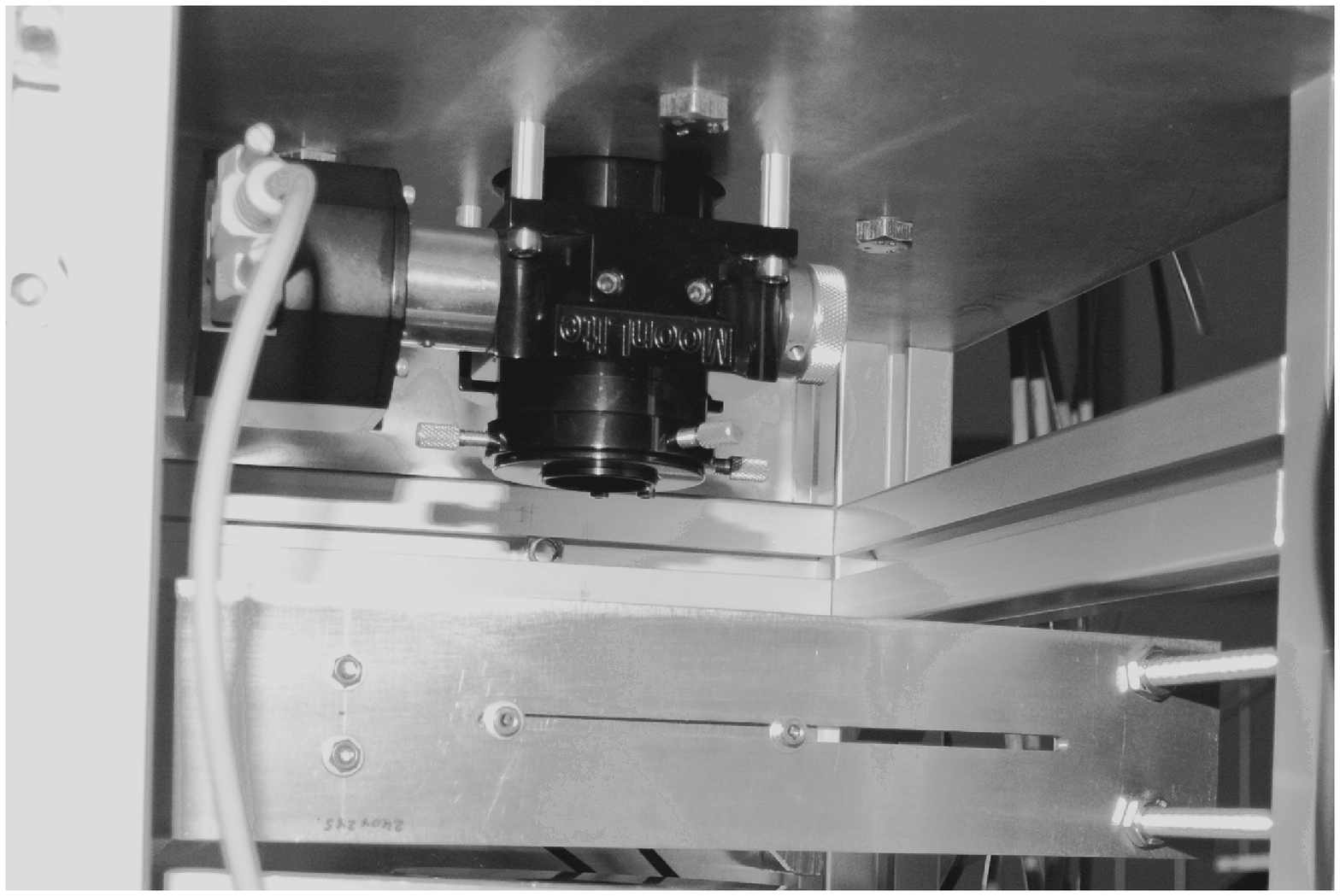} 
}
\caption{PISCO2: inside views. 
Left: general overview. The remote controlled
Risley prisms and the two corresponding USB-interface boxes are located
at the bottom of this image. 
Right: close up view of the motorized focusing system. 
}
\label{fig:pisco2-inside}
\end{figure*}

\begin{figure*}
\centerline{
\includegraphics*[height=4cm]{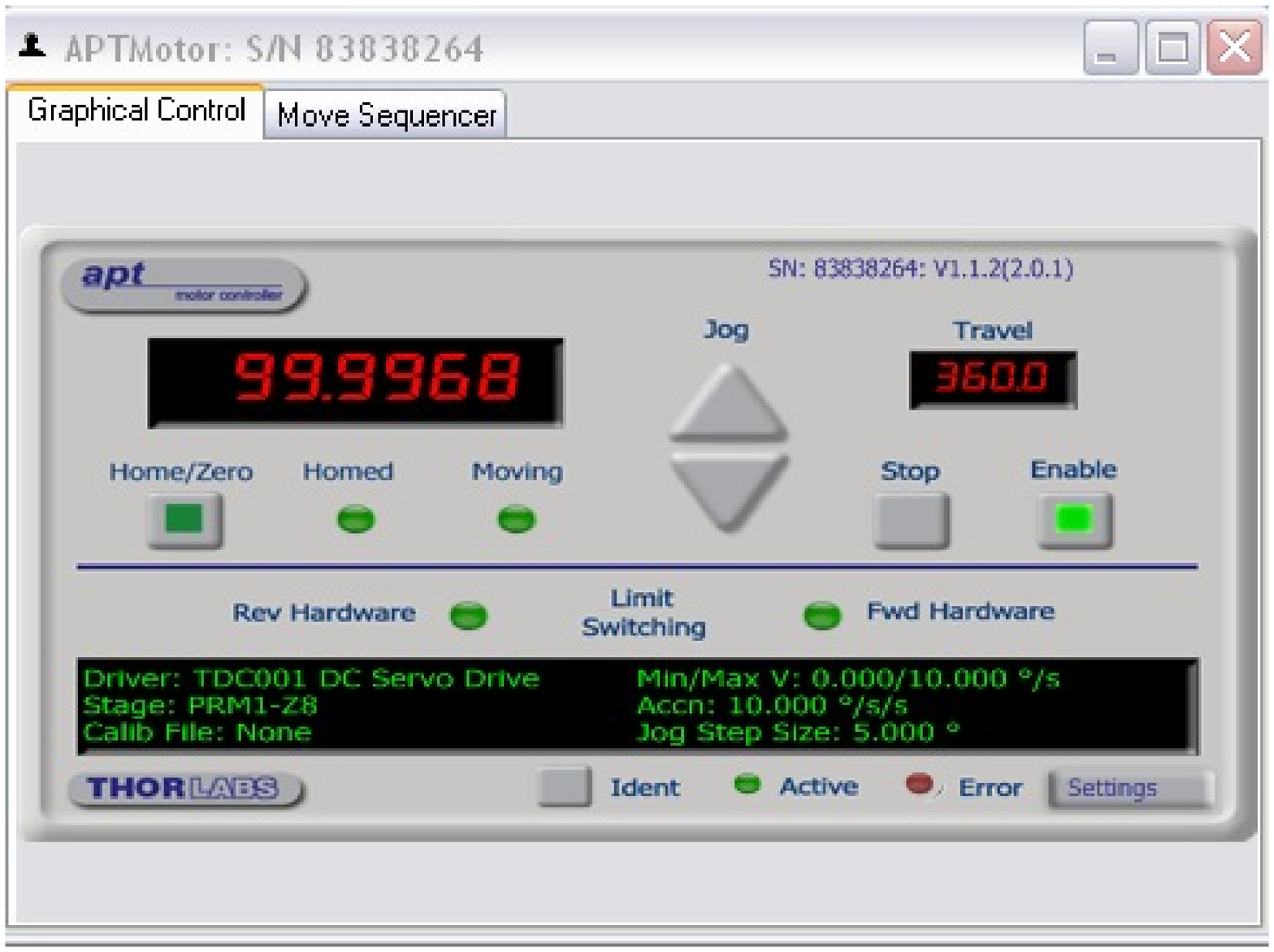}
\includegraphics*[height=4cm]{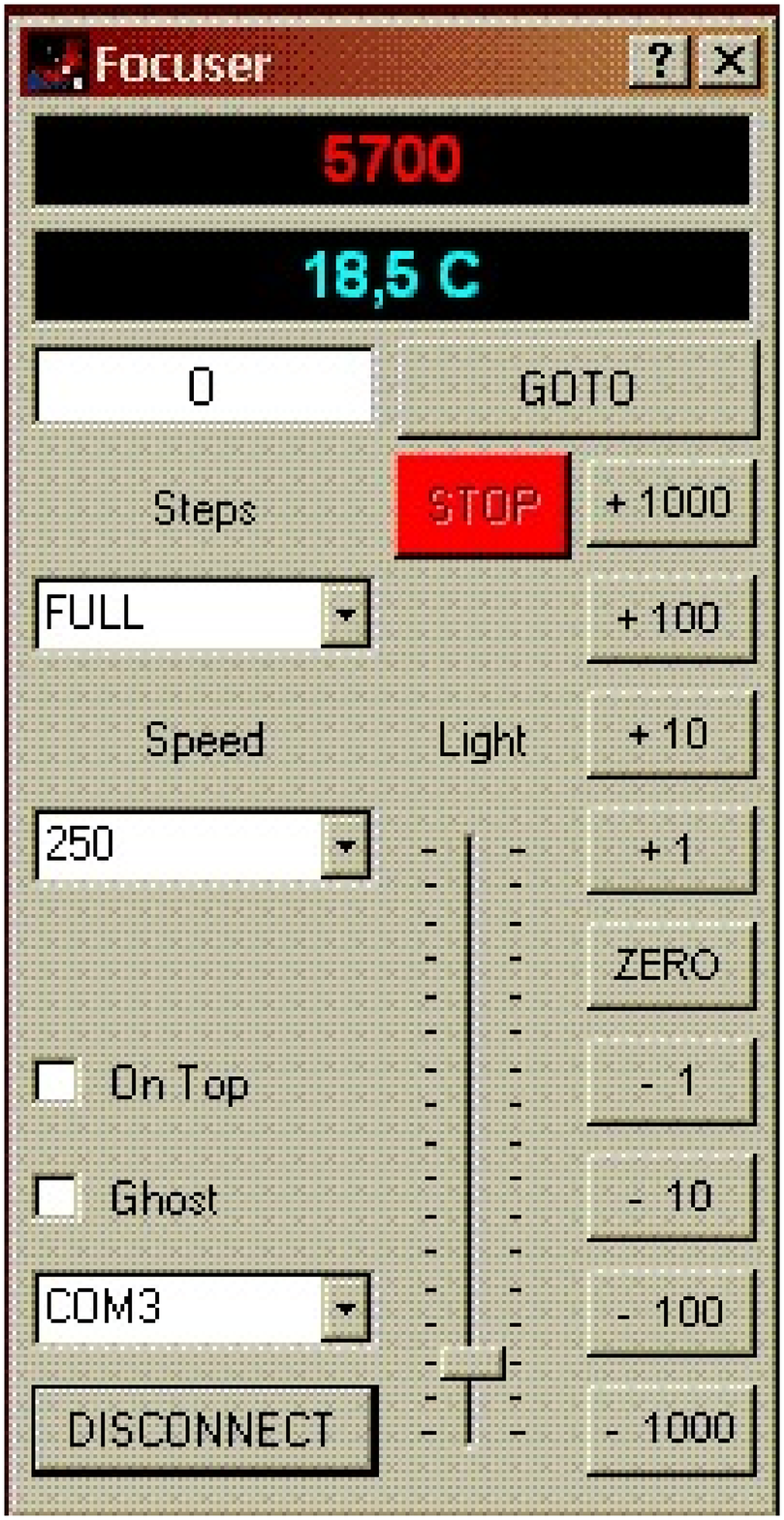}
\includegraphics*[height=4cm]{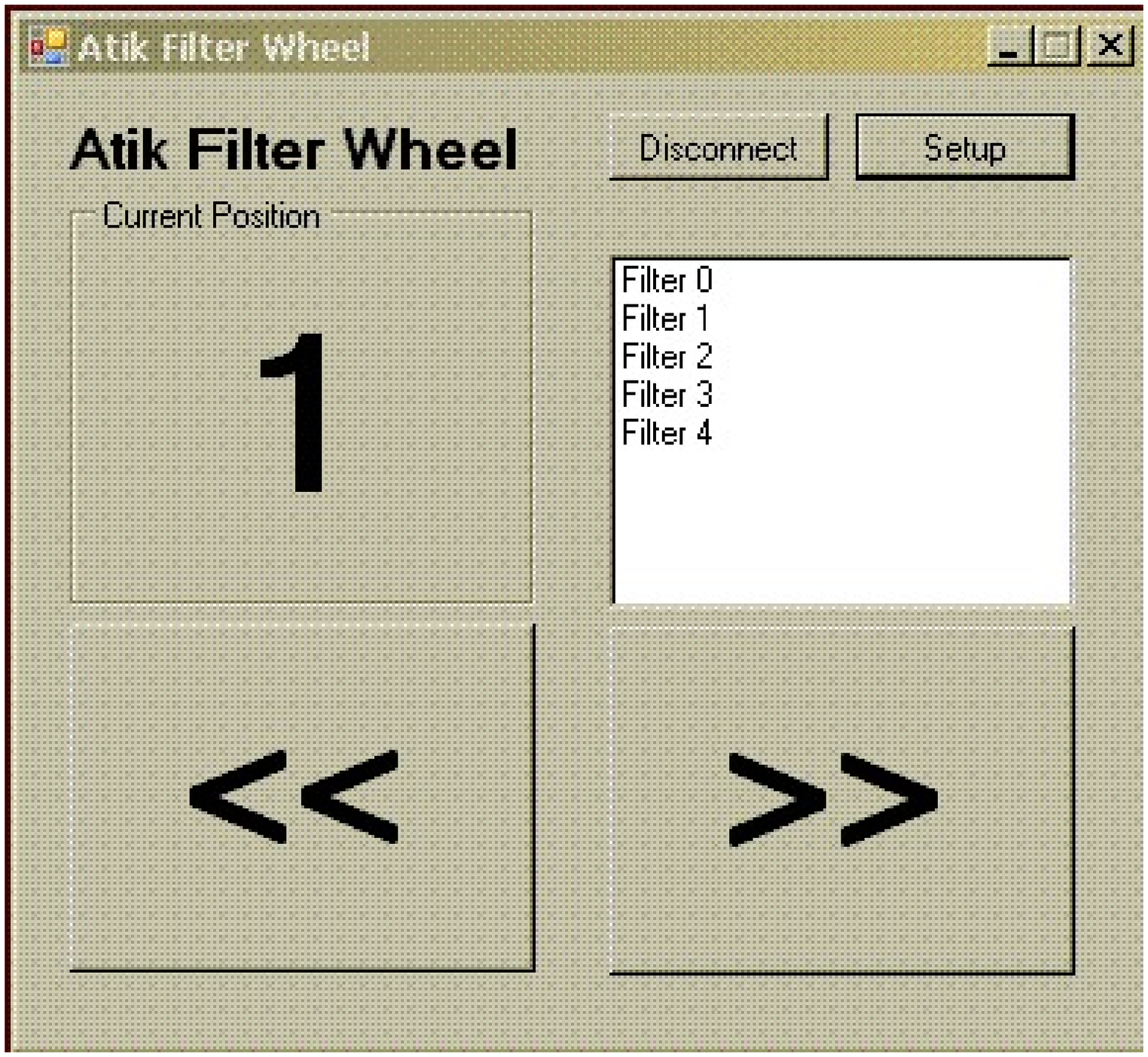}
}
\caption{PISCO2: remote control of 
the Risley prisms (left panel),
 motorized focusing (central panel) and filter wheel selection
 (right panel).}
\label{fig:remote-functions}
\end{figure*}

PISCO2 was entirely made at OCA
 between 2010 and 2012 (see Figs.~\ref{fig:pisco2-L76},
\ref{fig:pisco2-inside}).
Most of the mechanical parts were machined in the OCA workshops.
When possible 
complete manufactured units were integrated into PISCO2, 
to reduce the development time and the cost. For example, 
this was the case for the remote-control units 
dedicated to the motorized focusing or the 
angular control of the Risley prisms (see Fig.~\ref{fig:remote-functions}).

\subsection{Optical design}
\label{sec:pisco2-optical-layout}

\begin{figure*}
\centerline{\includegraphics*[width=14cm]{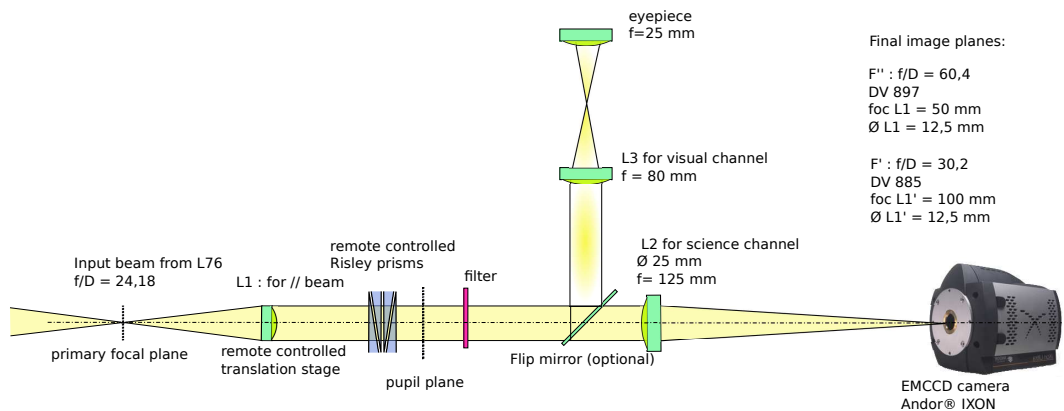} }
\caption{PISCO2: optical layout.}
\label{fig:optical-layout}
\end{figure*}

The optical layout 
(see Fig.~\ref{fig:optical-layout})
is similar to that of PISCO (see Prieur et al., 1998). 
An achromatic lens L1 of
focal length $F_1$ is placed at a distance $F_1$ from the
primary focus plane of the telescope. This lens thus creates
a collimated beam which is used for placing the Risley prisms
and the filters. A second achromatic lens L2 focuses this parallel
beam onto the image plane of the detector. 
A retractable mirror and a set of two lenses (L3 and eyepiece)
enable a visual inspection of the images.
The Risley prisms are located very close to the pupil plane.
Contrarily to PISCO, there is no  
field lens in the focal plane of the telescope. This improves
the overall transmission, but imposes a longer distance 
for the collimated beam.

PISCO2 can use different detectors, which may have different
pixel sizes. The adaptation can be done by changing  
the lens L1, and choosing its focal length value accordingly.
For instance, two L1 lenses of 50 and 100~mm focal length have been used 
for the ANDOR DV897 and~DV885 detectors, respectively. 

The available filters are: 
\begin{itemize}
\item{the IRC (IR-cut), which is a lower-pass
filter rejecting all wavelengths above 700~nm;
}
\item{the Schott BG39, which is a band-pass between 350 and 600~nm;}
\item{and the AF (anti-fringe or V-block), which is a bandpass
450--650~nm;}
\end{itemize}

The AF filter considerably reduces 
the secondary spectrum of the 76-cm refractor,
with no significant loss of energy in the $V$ band. When combined 
with the transmission of the lenses and 
the quantum efficiency response of the detector, the resulting
transmission curve of those filters is close to a standard
$V$ filter with a maximum around 570~nm. 

Most optical settings of PISCO2 can be remotely 
controlled with a computer:
\begin{itemize}
\item{Focusing} 
\item{Filter selection}
\item{Risley prism correction}
\end{itemize}

Motorised focusing and filter wheel selection
are controlled by ASCOM interface programs provided
by the corresponding manufacturers. Focusing is driven by a step motor
whose step corresponds to a translation of 4$\mu$m, with a full range 
of 40~mm.  Risley prisms can be set 
by the THORLABS program which works with a USB interface
 (see Fig.~\ref{fig:remote-functions}).

\subsection{Description of the detectors}
\label{sec:detectors}

The PISCO2 instrument was successively fitted with
two EMCCD detectors from ANDOR Technology:
an iXON DV885 and an iXON DV897 (see ANDOR, 2014). 
We first used the DV885 belonging to RG, 
that is equipped with a front-illuminated EMCCD chip. 
We then used the DV897 belonging to F.~Vakili (FV),  
which is more recent and has better performances.
It has a back-illuminated EMCCD with a higher
quantum efficiency and is fitted with a more elaborated cooling
system. Another advantage of the DV897 is 
its higher reading frequency. 


The main characteristics of the two detectors 
are given in Table~\ref{tab:detectors}. 
For each one, we indicate its full format in pixels (Col.~2),
the pixel size in $\mu$m (Col.~3), the digitization depth in Col.~4, 
i.e. the number of bits per pixels
used for encoding the output values, the maximum frequency rate (Col.~5) used
for reading out the pixel data, the theoretical quantum efficiency of the   
detector (Col.~6), and the cooling temperature that was used during our 
observations (Col.~7).

The quantum efficiency values in Col.~6 of Table~\ref{tab:detectors}
are those given by
the EMCCD chip constructor for wavelengths in the range 550--720~nm. 
The overall effective efficiency is unfortunately much smaller.
With PISCO in Merate (Italy), 
we have done some comparative tests in 2011 with 
the iXON DV885 and the PISCO ICCD cameras (see Scardia et al., 2013). 
For speckle observations, the DV885 was roughly equivalent to the 
ICCD. Note that the ICCD camera is fitted 
with a R-photo-cathode amplifier whose quantum efficiency is 7\% only (!).

\begin{table*}
\caption{Main characteristics of the ANDOR iXON EMCCD cameras 
used for our observations. The quantum efficiency in Col.~6 
is the value given by the manufacturer for the EMCCD chip.
The overall effective efficiency is much smaller.}
\label{tab:detectors}
\begin{tabular}{lcccccccl}
\hline
 & & & & & & & & \\
Name & Format & Pixel size & Digitization & Read freq.
& Quantum Eff. & Max. cooling T. \\
 & (pixels) & ($\mu$m) & (bits) & (MHz) & (\%) & ($^\circ$ C) & \\ 
 & & & & & & & \\
\hline
 & & & & & & & \\
DV885 & 1004 $\times$ 1002 & 8 $\times$ 8 & 14 & 27 & 60-65 & $-$70 \\
 & & & & & & & \\
DV897 & 512 $\times$ 512 & 16 $\times$ 16 & 14 & 10 & 80-92 & $-90$ \\  
 & & & & & & & \\
\hline
\end{tabular}
\end{table*}

\begin{figure*}
\centerline{
\includegraphics*[height=7cm]{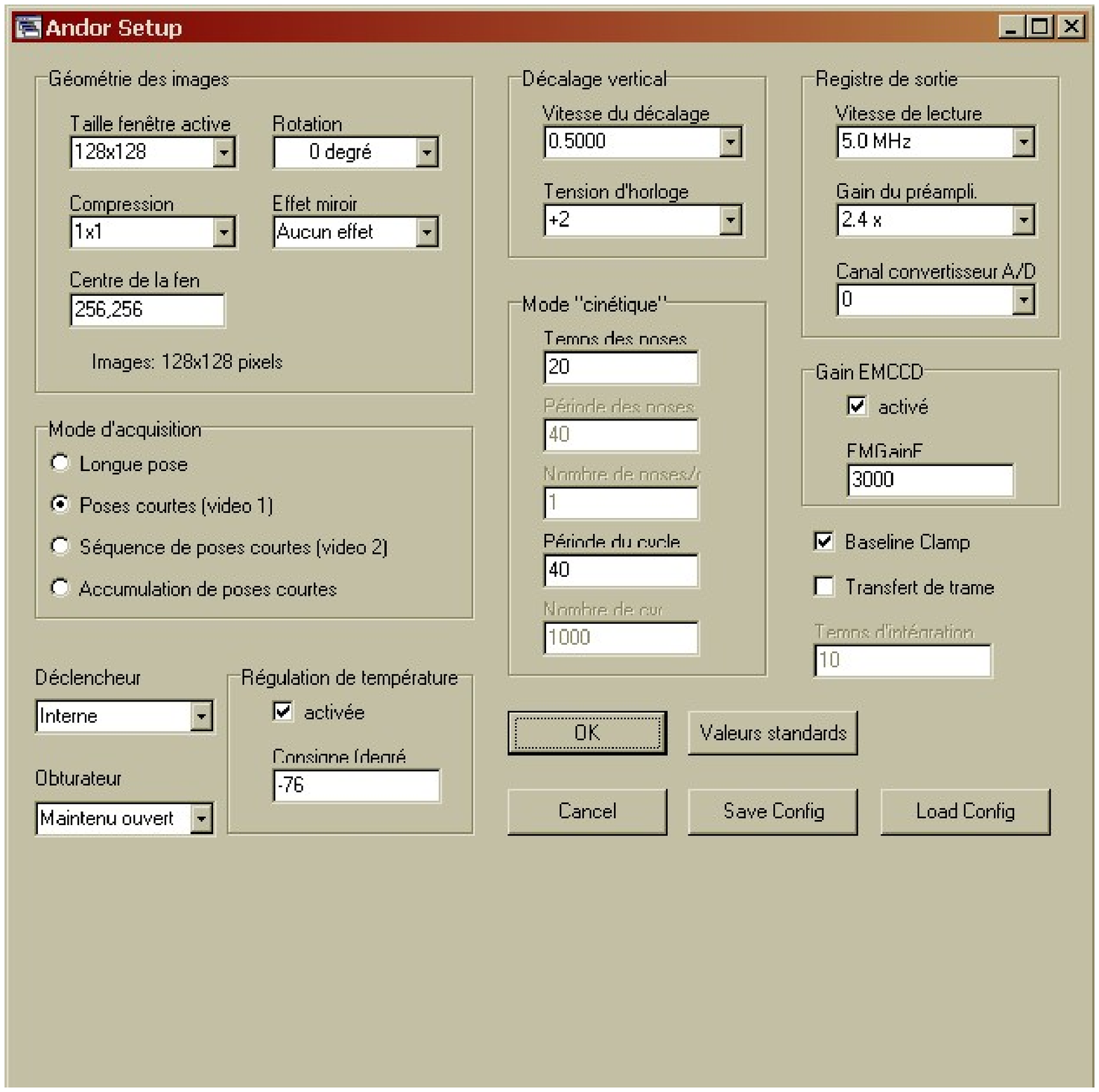}
\includegraphics*[height=7cm]{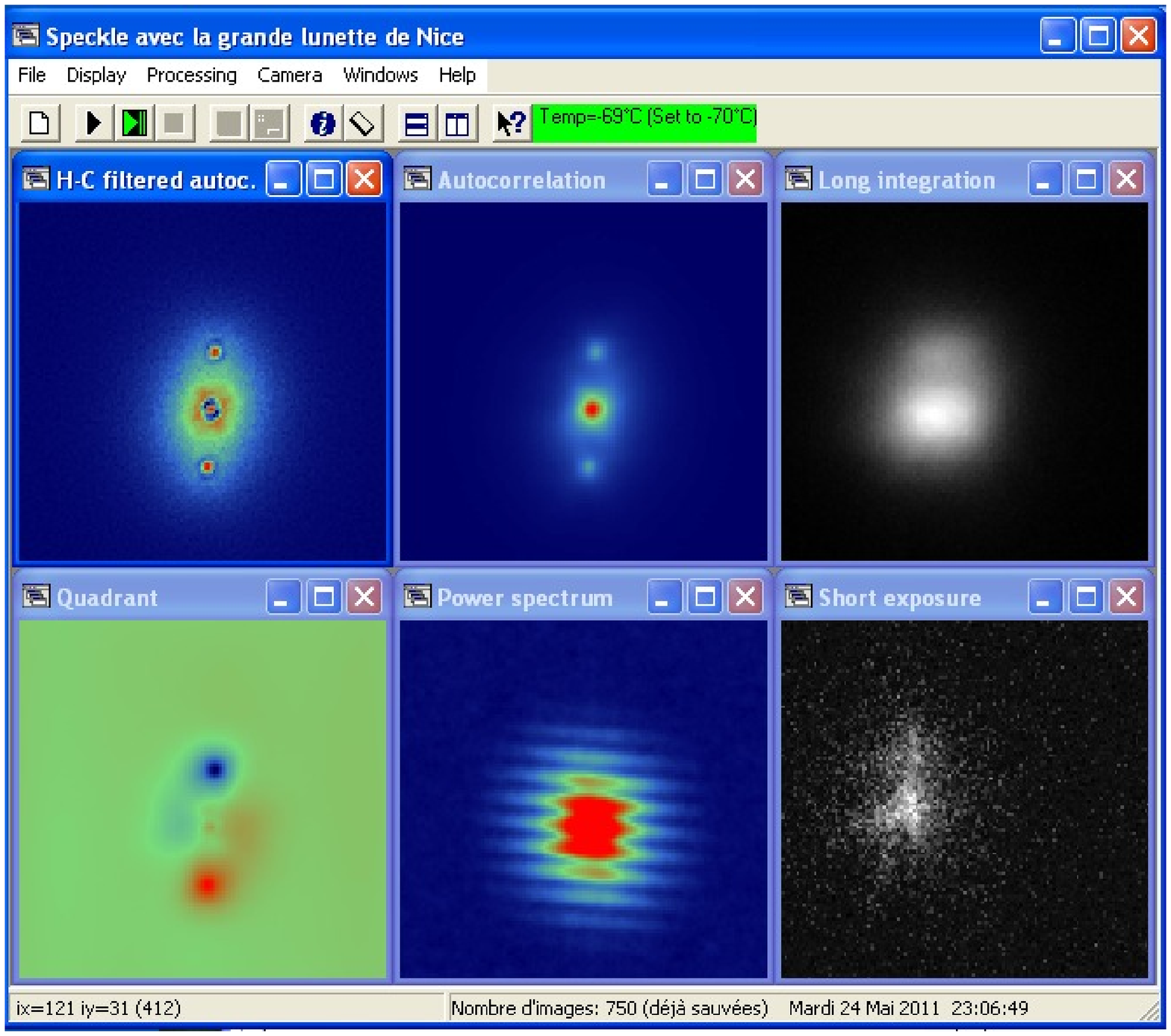}
}
\caption{Program \texttt{buildspeck1}
used for data acquisition with ANDOR cameras and real-time
processing. Left: camera setup; right: example of
image processing of the binary star BU~1273 ($\rho=1.\arcsec3$, $m_V=9.6$
$\Delta m_V = 1.3$).}
\label{fig:buildspeck1}
\end{figure*}

For both detectors, the image transfer is done through a dedicated link 
between the detector and a CCI-22 frame grabber board installed 
on a PCI slot of the acquisition computer, which allows
a reading frequency as high as 27~MHz.
This would correspond to an acquisition rate
of 27~full-size images per second. 
In practice, we used a lower readout frequency (5.13~MHz)
to reduce the noise in the images, 
Those detectors can be used in EM mode, which reduces the read-out-noise 
of the output register to less than one electron.

We wrote a specially designed program (\texttt{buildspeck1})
for handling the data acquisition 
of those two detectors (see Fig.~\ref{fig:buildspeck1}). 
This program controls the electronic settings
and the basic functions of the EMCCD detectors.
It also performs real-time processing of speckle interferometry observations.

Exposure times of elementary frames are set in the range 20--30 msec 
for speckle observations. The standard format of the acquisition window is
128$\times$128 pixels which corresponds to a field of view
of 9.5\arcsec$\times$9.5\arcsec\ for the DV897 camera. 
For faint objects or wide pairs,
a wider field of 256$\times$256 pixels on the detector
can be used with a binning factor of $2\times2$ which 
thus amounts to 128$\times$128 pixels for the elementary frames. 
The EM gain is generally set to values of about  170--200 and 3000
for the DV885 and the DV897, respectively.
To avoid saturation with bright objects, the EM gain can be 
reduced or even put to ``off'', which corresponds to observing
in conventional CCD mode.   

\section{Correction of the atmospheric dispersion with Risley prisms}
\label{sec:risley}

For ground-based observations of astronomical objects, 
the atmosphere behaves like a dispersive prism
(see f.i., Simon, 1966).
Polychromatic images are spread into a small vertical spectrum.
This effect can be neglected for observations close to the zenith,
but is very strong at low elevations, 
and can severely degrade the angular resolution. 
For instance in the visible, 
for a $\Delta \lambda=250$~nm bandpass centered at $\lambda=500$~nm, 
the typical atmospheric dispersion  is $\Delta\theta=1$\arcsec\ 
for an elevation $h = 60^\circ$
and $\Delta\theta=2$\arcsec\  for $h =30^\circ$.

\begin{figure}
\centerline{
\includegraphics*[height=3.cm]{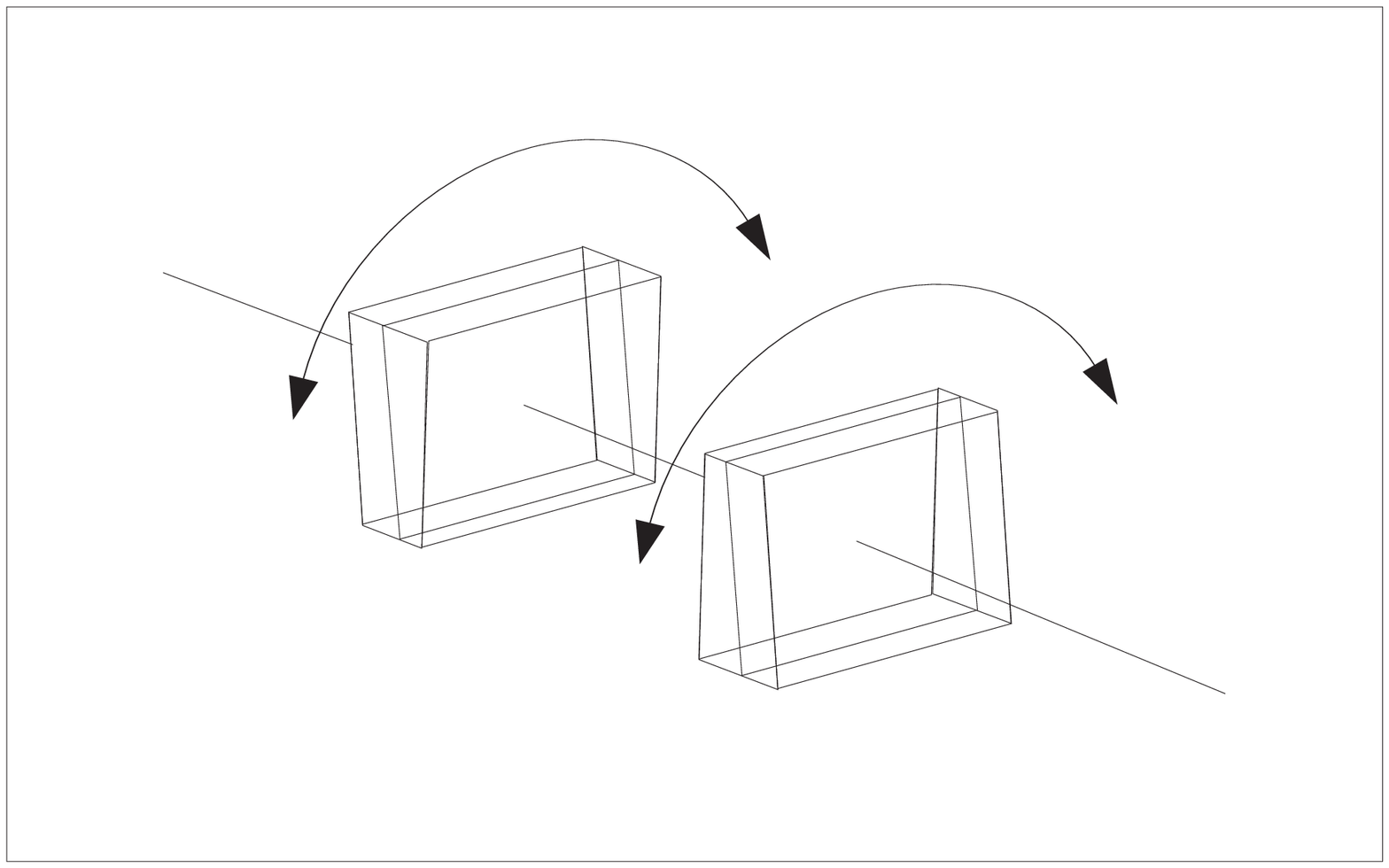}
\includegraphics*[height=3.4cm]{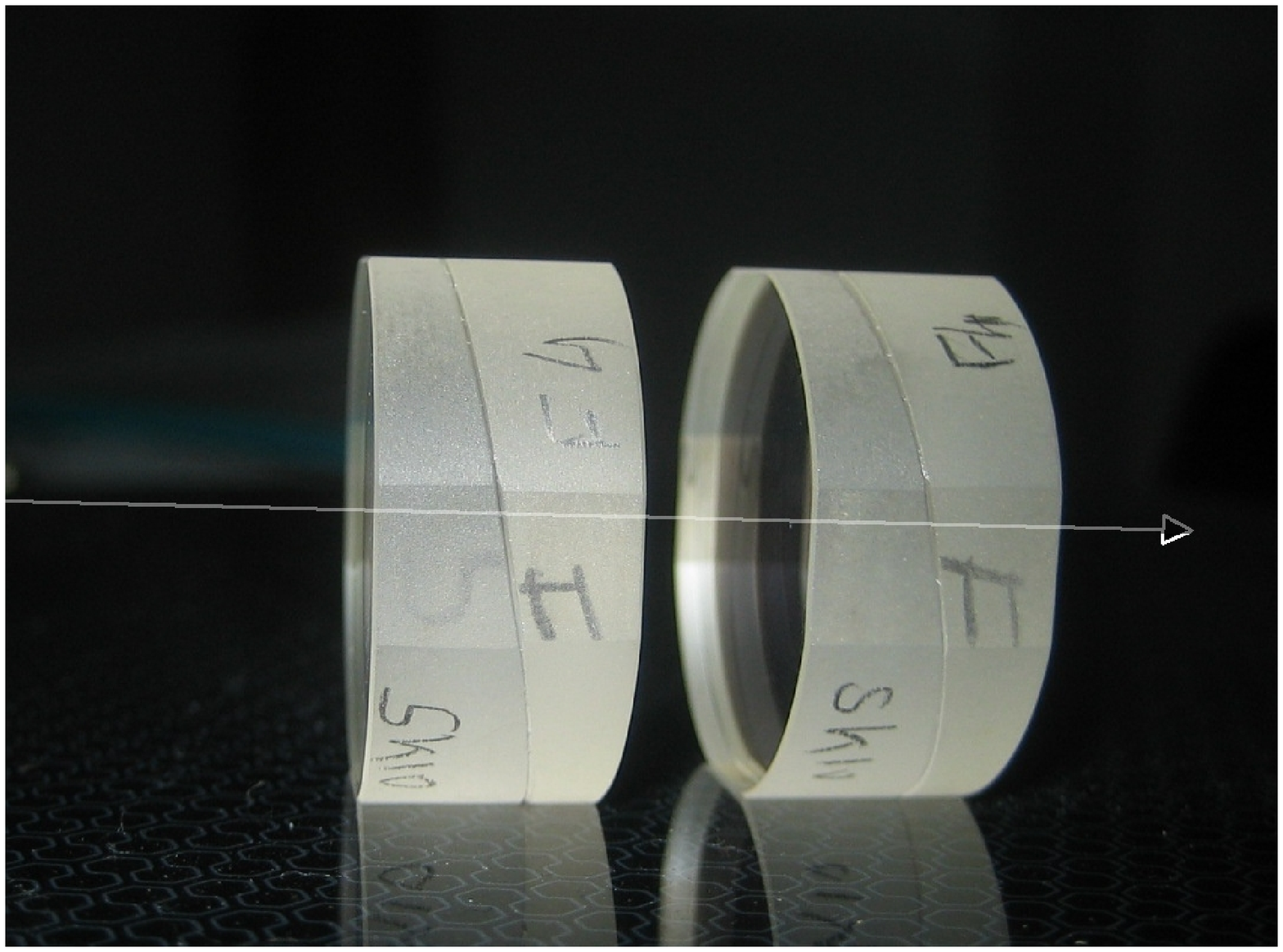}
\includegraphics*[height=3.4cm]{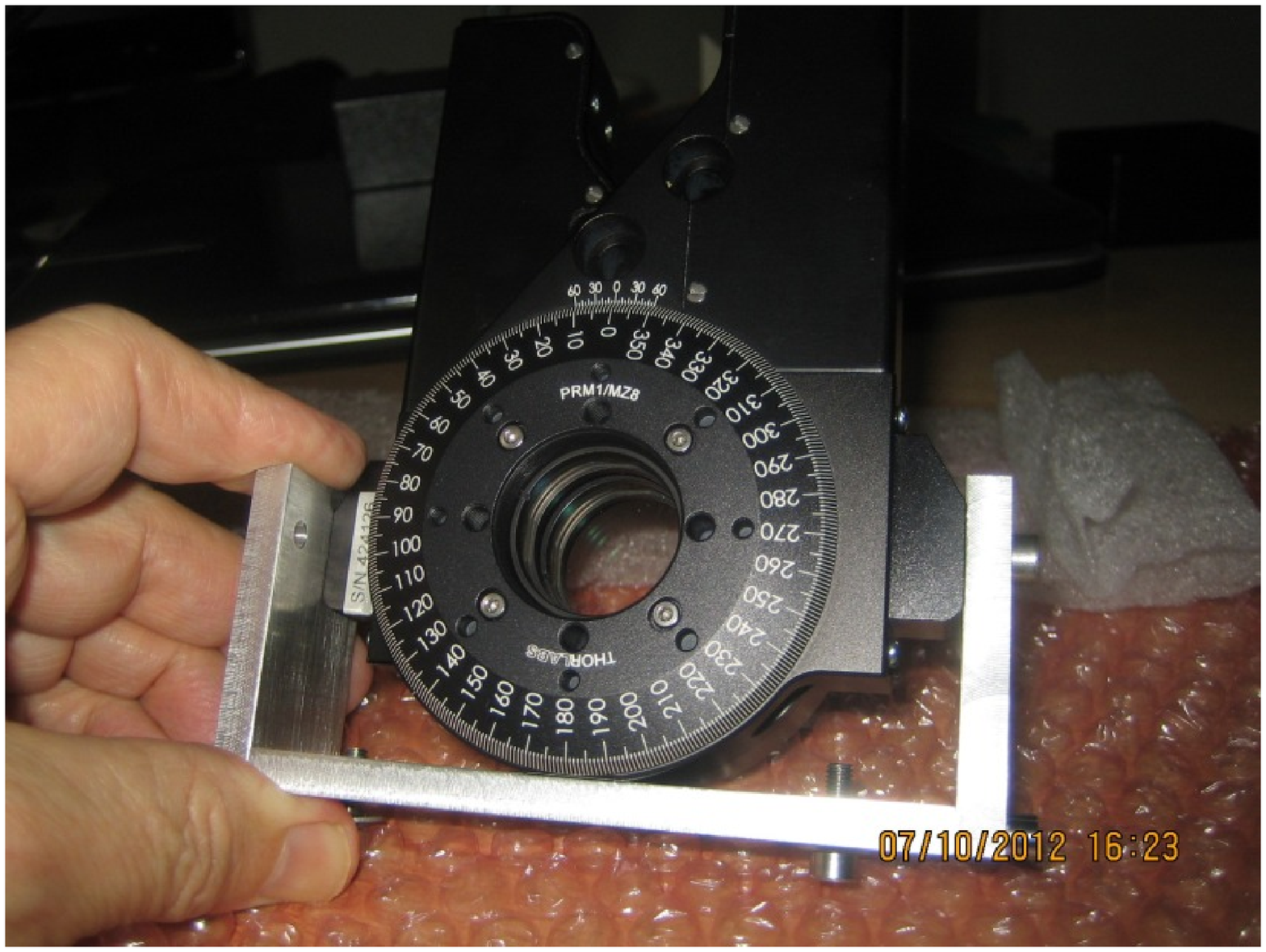}
}
\caption{PISCO2 Risley prisms used for atmospheric dispersion correction.
From left to right: schematic drawing, view of the prisms, and 
THORLABS motorized wheel}
\label{fig:risley-prisms}
\end{figure}

PISCO2 contains an atmospheric dispersion corrector to
circumvent this problem. 
This corrector is similar to that of PISCO (see Prieur et al., 1998) and
is based on ``Risley prisms''
(see f.i. Breckinridge {\it et al.} 1979, Walner \& Wetherell, 1990).
They consist in two identical sets of prisms
that can be rotated to produce a tunable chromatic dispersion both
in amplitude and direction.
Each set is made of two prisms that have different
indices and roof angles, and that are glued together in an upside-down
position (see Fig.~\ref{fig:risley-prisms}).

Like for PISCO, we have used the same combination of Schott glasses (F4, SK10)
which was also used for the Kitt Peak speckle camera
(Breckinridge {\it et al.}, 1979).
There are of course other possibilities, like the combinations
proposed by Wallner \& Wetherel (1990), which
are closer to the atmospheric dispersion curve.
The F4$+$SK10 combination has the advantage of a low cost 
and was sufficiently efficient for our purpose. 

The Risley prisms of PISCO2 have been designed to have
a null mean deviation, and a dispersion allowing
atmospheric correction from the zenith down to
an elevation of 30$^\circ$, when using the DV897 detector.
The F4 and SK10 prisms have a roof angle of 10.0$^\circ$ 
and 9.92$^\circ$ respectively. The tolerance on the roof angle is 6 arcmin,
and the surface accuracy is $\lambda/4$ (for $\lambda$=550~nm).
Those specifications lead to a reduction of the
residual dispersion down to a level
smaller than 0.01\arcsec\ for any object
located at an elevation larger than 30$^\circ$, with the
AF filter (450-650~nm bandpass). 
This value is much smaller than the diffraction limit
of 0.16\arcsec of the 76-cm refractor. 

\begin{figure*}
\centerline{
\includegraphics*[height=5cm]{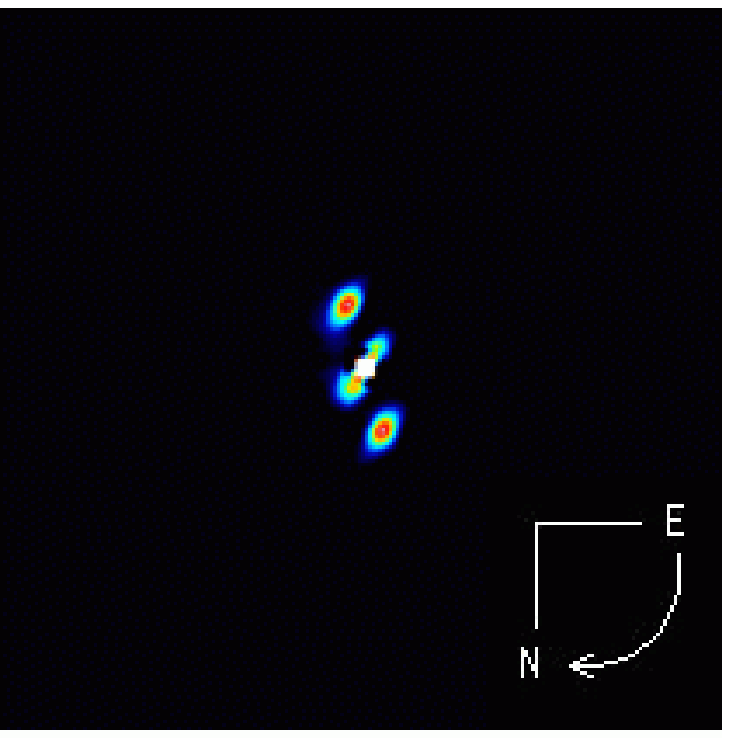} 
\includegraphics*[height=5cm]{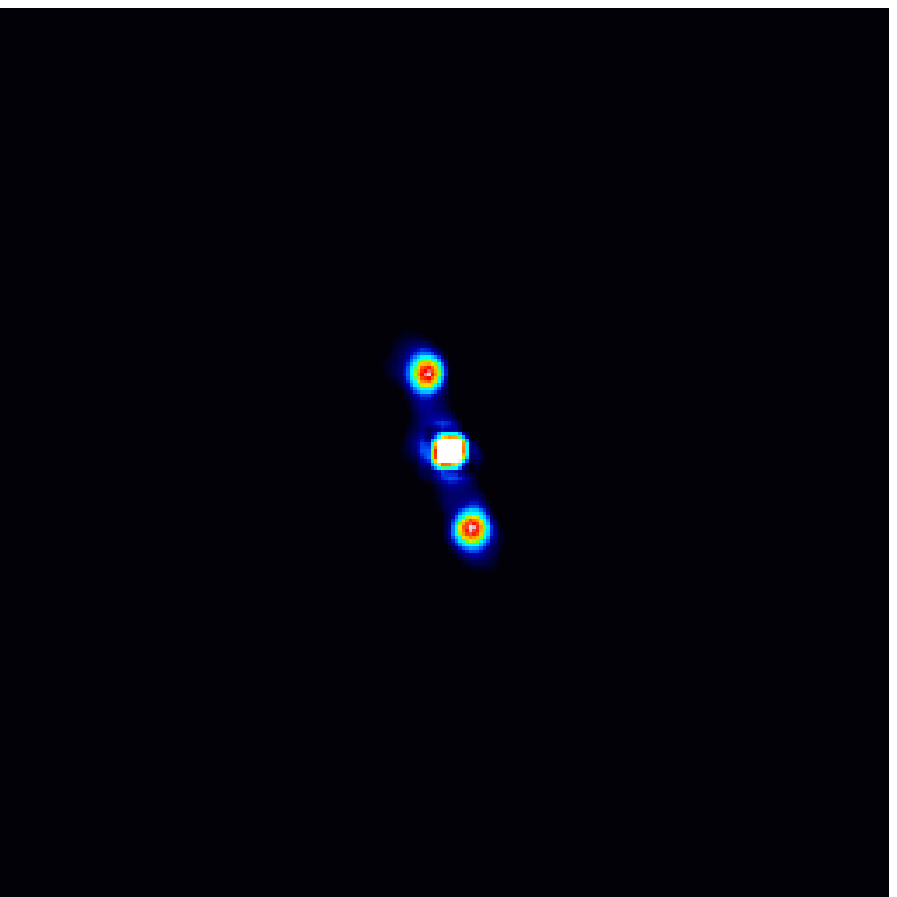} 
}
\caption{Effect of the 
atmospheric dispersion on the autocorrelation of a binary star (left) and 
correction with Risley prisms (right).
Object: 
A2724, 
$\rho = 0.\arcsec8$, $\theta=197^\circ$, $m_{V}=8.4$,
$\Delta m_{V}=0.7$, observed at a zenith distance of $z=42^\circ$.
}
\label{fig:risley-correction1}
\end{figure*}

\begin{figure*}
\centerline{
\includegraphics*[height=5cm]{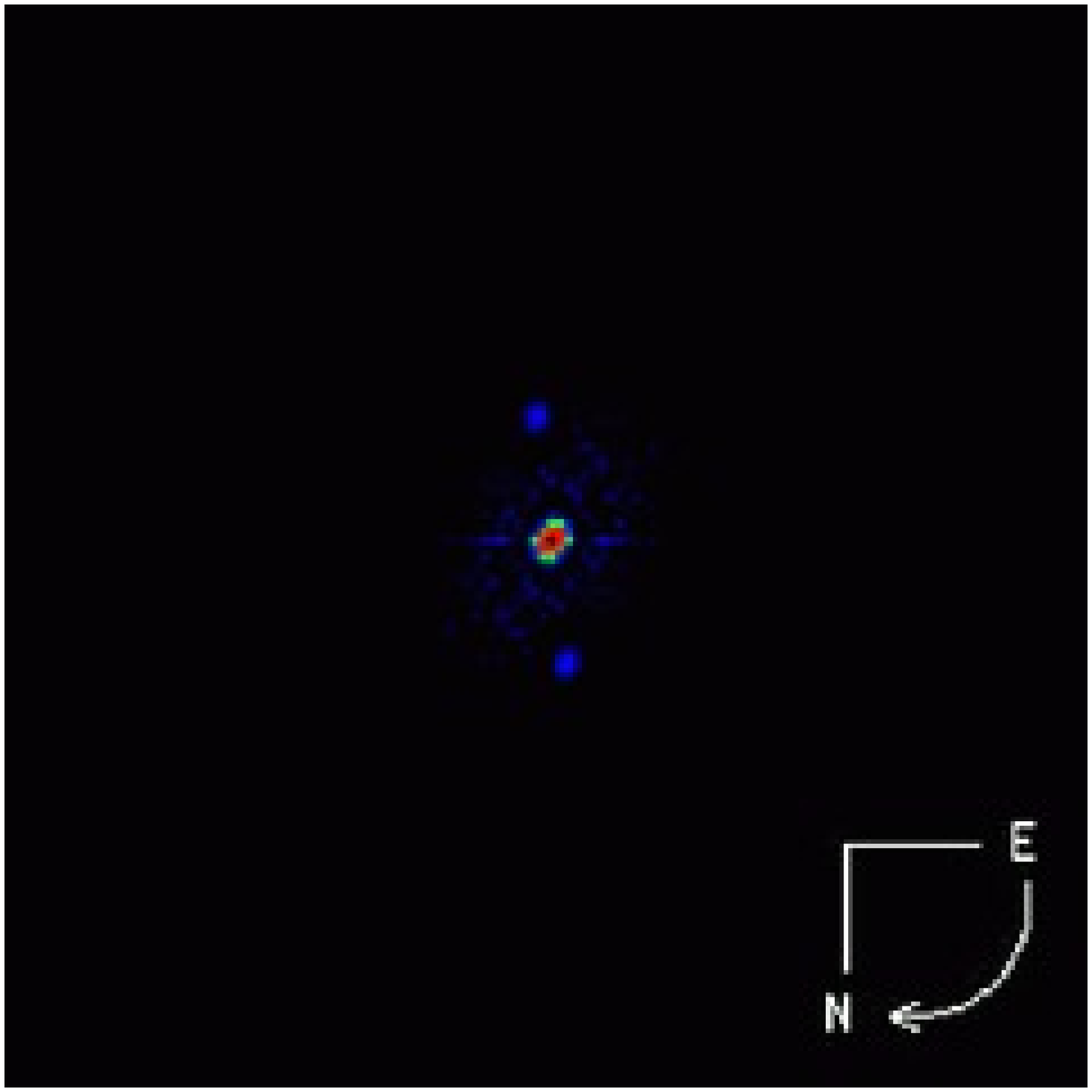} 
\includegraphics*[height=5cm]{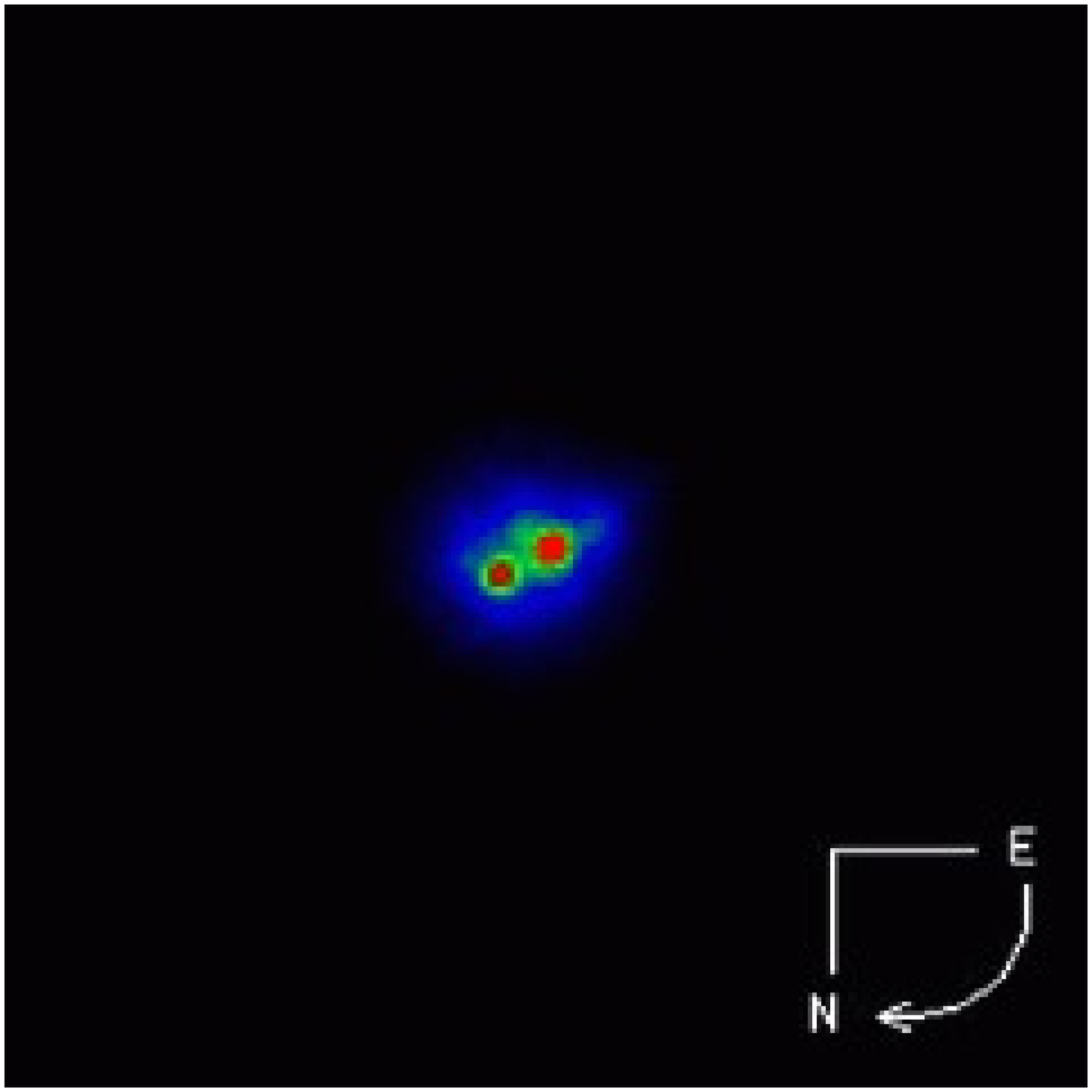} 
}
\caption{Left: mean autocorrelation of RST 4712
($m_{V}=9.8$, $\Delta m_{V}=1.2$,
$\rho = 1.\arcsec1$, $\theta=193^\circ$).
Right: image obtained by adding recentered
selected images (Lucky imaging) of 
STF 2696 
($m_{V}=7.9$, $\Delta m_{V}=0.8$,
$\rho = 0.\arcsec5$, $\theta=298^\circ$). 
Due to the large zenith distance of those
observations (50$^\circ$ and 40$^\circ$, respectively) 
Risley 
correction was needed (and successfully applied) in both cases. 
}
\label{fig:risley-correction2}
\end{figure*}

During the observations, a specially designed program  
(\texttt{pisco2\_risley.cpp})
computes the elevation of the star and the atmospheric
dispersion inferred from models
(Owens, 1967, formulae 29--31).
The Risley prisms are then rotated with the THORLABS
remote-control program, so that
their total dispersion has the same magnitude as
the atmospheric dispersion and an opposite direction.
The observations have shown that the
correction is very good, even for objects whose declination
is as low as -7$^\circ$, which is the pointing lower limit 
for the present instrumentation
setup, due to the camera cable length.
(see Figs.~\ref{fig:risley-correction1},
\ref{fig:risley-correction2}).

\section{Scale calibration with a grating mask}
\label{sec:calibration}

The on-sky pixel scale of the whole instrument (telescope and PISCO2)
 can be calibrated in an absolute manner with a grating mask
placed in front of the refractor objective lens. This is the method 
used for calibrating PISCO in Merate, as described in
detail in Scardia et al. (2007).
We present here the procedure we followed in September 2012 for calibrating
the scale of the DV897 detector. 

\begin{figure*}
\centerline{
\includegraphics*[height=6cm]{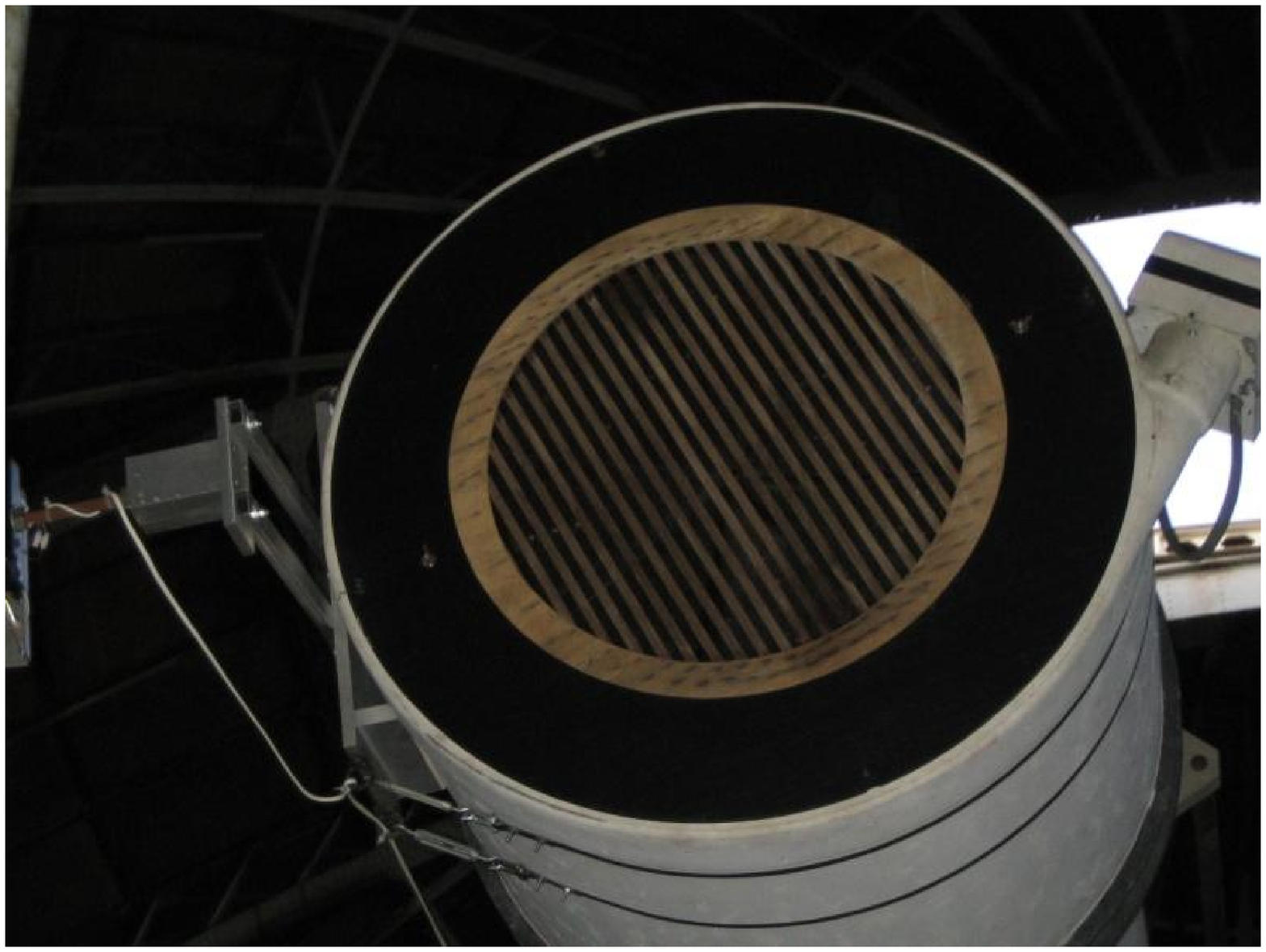} 
\includegraphics*[height=6cm]{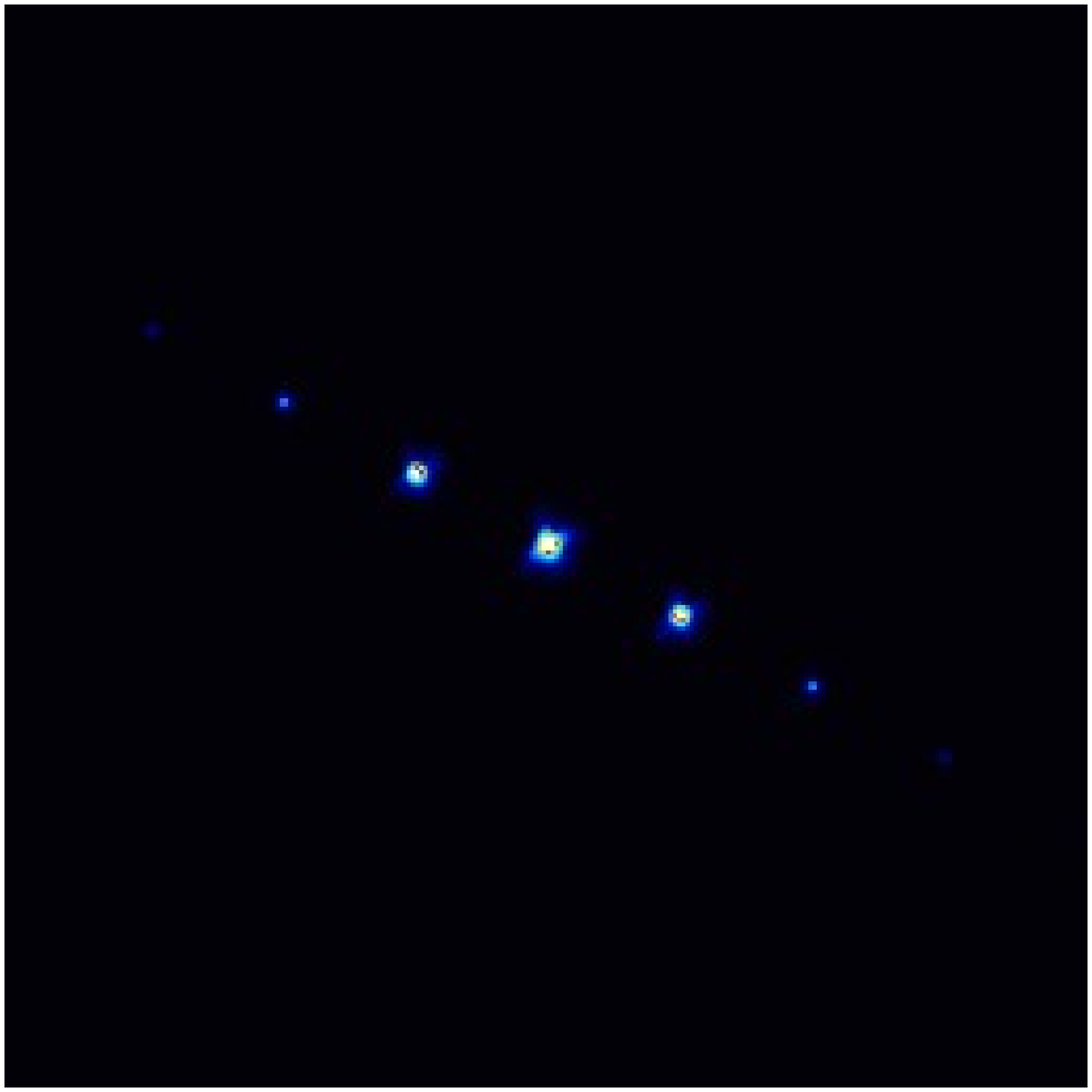} 
}
\caption{Scale calibration with a grating mask. Left: mask placed on top
of the 76cm-refractor. Right: mean autocorrelation of the diffraction
pattern obtained with $\epsilon$ Leo. 
}
\label{fig:calibration}
\end{figure*}

As shown in Fig.~\ref{fig:calibration}, 
a grating mask was mounted on the entrance baffle of the telescope.
This mask was originally designed 
for the calibration of the OCA 50-cm refractor. 
It has a diameter of 50~cm and
a period of 22.43$\pm$0.10~mm.
To obtain measurements with a high precision, we 
used a narrow-band filter, centered on 
$\lambda = 570$~nm, with a bandwidth of $\Delta \lambda = 10$~nm.
The corresponding period of the diffraction pattern was then
$\lambda / p = 5\arcsec.241$. For this
calibration, we observed the single star $\epsilon$~Leo.
The measurements were made on the mean of a series 
of a few 1000-image data cubes.
The equivalent focal length with PISCO2 and
the DV897 detector was found to be $F = 44.66 \pm 0.2$~m
and the scale of 0.0739 \arcsec/pixel. 

The 76-cm refractor has actually a free aperture of 74~cm.
The corresponding limit of diffraction $\rho_D = \lambda/D$ is 0\arcsec.16 with
$\lambda = 570$~nm.
The sampling of this detector mounted on PISCO2 is thus smaller than 
$\rho_D/2$ for $\lambda = 570$~nm. This sampling is
compatible with the Nyquist-Shannon criterion,
 and therefore allows measurements down to the telescope diffraction limit.

The calibration of the origin of the position angles was
done recording star trails caused by the diurnal motion. We used the largest
available field for this purpose, which was $53\arcsec \times 40\arcsec$
with the DV897.

\section{Conclusion}
\label{sec:conclusion}

The first observations have shown that PISCO2 performances
have reached the initial goal. 
Fitted with remote-control components such as a filter wheel,
a motorized focusing system, and an atmospheric dispersion corrector,
this instrument is very easy to operate.

PISCO2 is dedicated to the observation of visual binary stars.
In the last few years, it has already permitted
to obtain numerous measurements, with a average rate of about 2100 objects/yr.
With the DV897 detector, we have been able to observe
faint close binary stars
with angular separations as small as 0\arcsec.16, and visual magnitudes
of about~16. We also have measured some particularly difficult
systems with a magnitude difference between the two components
of about~4~mag. 

Those performances are very promising for the detection and study 
of large sets of yet unknown (or partly measured) binaries 
with close separation and/or large magnitude difference, 
many of which could not be measured by space missions. 
PISCO2 may thus provide a significant contribution to stellar physics.

\bigskip
\noindent
{\bf Acknowledgements}

We are indebted to the direction of Observatoire de la C\^ote
d'Azur for 
allowing us to use the 76-cm refractor.
We thank the workshop staff of this observatory
for their technical support
and Y.~Bresson (OCA) 
for his contribution to the optical design of PISCO2.
We are also very grateful to R.W.~Argyle (Cambridge, U.K.) for
checking the English of this paper.

\bigskip
\def\fullref#1{\textbf{#1}\\}



\begin{thebibliography}{myexample99}

\bibitem[ANDOR(2014)]{andor14}
ANDOR: 2014, ANDOR technology
\hbox{http://www.andor-tech.com}

\bibitem[Breckinridge at al.(1979)]{breckindge79} 
\fullref{Kitt Peak speckle camera}
Breckinridge, J.B., McAlister, H.A., Robinson, W.G., 1979,
Appl. Optics, 18, 7, 1034--1041.

\bibitem[Couteau (1970)]{couteau70}
\fullref{La grande lunette de l'Observatoire de Nice}
Couteau, P., 1970, L'Astronomie, 84, 213 

\bibitem[Couteau (1993)]{couteau93}
\fullref{Catalogue de 2700 \'etoiles doubles}
Couteau, P., 1993, Obs. C\^ote d'Azur, Nice,  2nd Edition.

\bibitem[Couteau \& Gili (1994)]{gili94}
\fullref{Mesures d'\'etoiles doubles faites \`a Nice, \'etoiles 
doubles nouvelles (24\`eme s\'erie) d\'ecouvertes \`a Nice.}
Couteau, P., Gili, R., 1994, A\&AS, 106, 377 

\bibitem[Couteau \& Gili (1997)]{gili97}
\fullref{Mesures d'\'etoiles doubles faites aux lunettes de 74 et 50 cm 
de l'Observatoire de la C\^ote d'Azur.}
Gili, R., Couteau, P., 1997, A\&AS, 126, 1 

\bibitem[Gili \& Bonneau (2001)]{gili01}
\fullref{CCD measurements of visual double stars made with 
the 74 cm and 50 cm refractors of the Nice Observatory (2nd series)}
Gili, R., Bonneau, D., 2001, A\&A, 378, 954

\bibitem[Gili \& Agati(2009)]{gili09}
\fullref{Measurements of double stars with the 76 cm refractor 
of the C\^ ote d'Azur 
observatory with CCD and emCCD cameras. 2nd part.}
Gili, R., Agati, J.-L., 2009, Observations \& Travaux, 74, 14

\bibitem[Gili \& Prieur(2012)]{gili12}
\fullref{Relative astrometric and photometric measurements
of visual binaries made with the Nice 76-cm refractor in 2008}
Gili, R., Prieur, J.-L., 2012, Astron. Nach., 333, 727--735

\bibitem[Labeyrie(1970)]{labeyrie70}
\fullref{Attainment of diffraction limited resolution in large telescopes by
Fourier analysing speckle patterns in star images}
Labeyrie A.: 1970, A\&A, 6, 85

%

\bibitem[Morlet et~al.(1999)]{morlet99}
\fullref{CCD measurements of visual double stars 
made with the 50 cm refractor of the Nice Observatory (2nd series)}
Morlet, G., Salaman, M., Gili, R., 1999, A\&AS, 135, 499

\bibitem[Morlet et~al.(2002)]{morlet02}
\fullref{Nice Observatory CCD measurements of visual double stars (4th series)}
Morlet, G., Salaman, M., Gili, R., 2002, A\&A, 396, 933 

\bibitem[Owens(1967)]{owens67} 
\fullref{Optical refractive index of air: dependence on
pressure, temperature and composition}
Owens,~J.C., 1967,
Applied Optics, Vol.~6, N$^\circ$1., 51--59.

\bibitem[Prieur et~al.(1998)]{prieur98}
\fullref{The PISCO speckle camera at Pic du Midi Observatory}
Prieur, J.-L, Koechlin, L., Andr\'e, C., Gallou, G., Lucuix, C.:
1998, Experimental Astronomy, vol 8, Issue 4, 297


\bibitem[Salaman et~al.(1999)]{salaman99}
\fullref{CCD measurements of visual double stars 
made with the 50 cm refractor of the Nice Observatory}
Salaman, M., Morlet, G., Gili, R., 1999, A\&AS, 135, 499


\bibitem[Scardia et~al.(2007)]{scardia07}
\fullref{Speckle observations with PISCO in Merate.
III.~Astrometric measurements of visual binaries in 2005
and scale calibration with a grating mask}
Scardia M., Prieur J.-L., Pansecchi L., Argyle R.W., Basso S., Sala M.,
Ghigo M., Koechlin L., Aristidi E., 2007, MNRAS, 374, 965--978


\bibitem[Scardia et~al.(2013)]{scardia13}
\fullref{Speckle observations with PISCO in Merate.
XII. Astrometric measurements of visual binaries in 2011}
Scardia, M., Prieur, J.-L., M., Pansecchi, L., Argyle, R.W., Span\'o, P.,
Riva, M., Landoni, M., 2013, MNRAS, 434, 2803--2813 

\bibitem[Simon(1966)]{simon66} 
\fullref{A practical solution of the atmospheric dispersion problem,}
Simon, G.W., 1966,
Astron.~J., {\bf 71}, 190.

\bibitem[Wallner \& Wetherell(1990)]{walner90} 
Wallner, E.P., Wetherell, W.B., 1990,
{\sl Broad spectral bandpass atmospheric dispersion correctors,}
Rapport technique Itek.

\end{thebibliography}
\end{document}